\documentclass[fleqn,usenatbib]{mnras}
\usepackage{newtxtext,newtxmath}
\usepackage[T1]{fontenc}
\usepackage{ae,aecompl}

\usepackage{graphicx}
\usepackage{amsmath}	
\usepackage{siunitx}
\usepackage{caption}

\usepackage{color}
\usepackage{soul}
\usepackage[dvipsnames]{xcolor}

\definecolor{vero}{rgb}{0.5, 0.3, 0.7}
\definecolor{christi}{rgb}{0.0, 0.58, 0.71}
\definecolor{alexdu}{rgb}{0.00, 0.50, 0.00}
\definecolor{matt}{rgb}{0.8, 0.0, 0.0}

\newcommand{\sifourfull}{Si~\textsc{iv} $\lambda\lambda$1393, 1402}
\newcommand{\cfourfull}{C~\textsc{iv} $\lambda\lambda$1548, 1550}
\newcommand{\nfivefull}{N~\textsc{v} $\lambda\lambda$1238, 1242}
\newcommand{\sifour}{Si~\textsc{iv}}
\newcommand{\cfour}{C~\textsc{iv}}
\newcommand{\nfive}{N~\textsc{v}}

\newcommand{\stokesv}{Stokes $V$}

\newcommand{\bz}{\langle B_z \rangle}
\newcommand{\nz}{\langle N_z \rangle}

\newcommand{\kms}{km\,s$^{-1}$}

\newcommand{\esp}{ESPaDOnS}
\newcommand{\ksi}{$\xi^1$~CMa}

\title[Confirmation of $\xi^1$ CMa's ultra-slow rotation]{Confirmation of $\xi^1$ CMa's ultra-slow rotation: magnetic polarity reversal and a dramatic change in magnetospheric UV emission lines}

\author[C. Erba et al.]
{C. Erba$^{1}$\thanks{E-mail: cerba@udel.edu}, 
M. E. Shultz$^{1}$,
V. Petit$^{1}$, 
A.W. Fullerton$^{2}$,
H.F. Henrichs$^{3}$,
O. Kochukhov$^{4}$,
\newauthor
T. Rivinius$^{5}$,
G.A. Wade$^{6}$
\\
$^{1}$ Department of Physics and Astronomy, Bartol Research Institute, University of Delaware, Newark, DE 19716, USA \\
$^{2}$ Space Telescope Science Institute, Baltimore, MD 21218, USA \\
$^{3}$ Anton Pannekoek Institute for Astronomy, University of Amsterdam, Science Park 904, 1098 XH Amsterdam, Netherlands \\
$^{4}$ Department of Physics and Astronomy, Uppsala University, Box 516, 751 20, Uppsala, Sweden \\
$^{5}$ European Organisation for Astronomical Research in the Southern Hemisphere (ESO), Casilla, 19001, Santiago 19, Chile \\
$^{6}$ Department of Physics and Space Science, Royal Military College of Canada, PO Box 17000 Station Forces, Kingston, ON, Canada K7K 0C6
}

\date{Accepted 2021 May 14. Received 2021 May 14; in original form 2021 April 5}

\pubyear{2021}

\begin{document}
\label{firstpage}
\pagerange{\pageref{firstpage}--\pageref{lastpage}}
\maketitle

\begin{abstract}
The magnetic $\beta$ Cep pulsator \ksi~has the longest rotational period of any known magnetic B-type star. It is also the only magnetic B-type star with magnetospheric emission that is known to be modulated by both rotation and pulsation. We report here the first unambiguous detection of a negative longitudinal magnetic field in \ksi~($\bz=-87 \pm 2$ G in 2019 and $\bz=-207 \pm 3$ G in 2020), as well as the results of ongoing monitoring of the star's H$\alpha$ variability. We examine evidence for deviation from a purely dipolar topology. We also report a new {\it HST} UV spectrum of \ksi~obtained near magnetic null that is consistent with an equatorial view of the magnetosphere, as evidenced by its similarity to the UV spectrum of $\beta$ Cep obtained near maximum emission. The new UV spectrum of \ksi~provides additional evidence for the extremely long rotation period of this star via comparison to archival data. 
\end{abstract}




\section{Introduction}

Approximately 10\% of O- and B-type stars host large-scale, strong, surface magnetic fields with strengths on the order of 1 kG \citep{Morel2015,Wade2016,Grunhut2017,Petit2019} that have a significant effect on stellar evolution \citep{Petit2017,Keszthelyi2019}. By trapping wind plasma within closed magnetic loops, these magnetospheres reduce the net mass-loss rate from the stellar surface \citep{udDoula2002}, and contribute to rapid stellar angular momentum loss (magnetic braking; \citealt{udDoula2009}). In extreme cases, this can lead to an essentially non-rotating star (e.g. HD 108, with a rotation period of $\sim 55$ yr; \citealt{Naze2010,Shultz2017a}). 

Among this subpopulation of extremely slow rotators, the magnetic $\beta$ Cep pulsator \ksi~(HD 46328, B0.5 IV) stands out due to its remarkable characteristics. The magnetic detection, first reported by \citet{Hubrig2006} and later confirmed by \citet{Silvester2009}, identified the strongest surface magnetic field measured for any B0 star ($\sim 1$ kG). Indeed, the star's magnetic field has not yet been fully characterized, since a complete rotation period of \ksi~ has not yet been observed. To date, longitudinal magnetic field measurements of the star span approximately 20 yrs, with the earliest recorded observations from the year 2000 \citep{Shultz2017}. Using FORS2 and SOFIN spectropolarimetric data, \citet{Hubrig2011} and \citet{Jarvinen2018} claimed a much shorter rotation period of 2.17937 d. These findings were challenged by \citet{Shultz2017,Shultz2018}, who showed that only an extraordinarily long rotation period of $\sim$ 30 yr was consistent with the high-resolution spectropolarimetric datasets obtained with \esp. \ksi~therefore exhibits the longest rotation period of any known magnetic B-type star \citep{Shultz2017,Shultz2018a}.

\ksi~also shows evidence for departures from a purely dipolar magnetic topology. \citet{Shultz2018} reported a crossover signature at magnetic null which, given the negligible $v \sin i$ implied by the extremely long rotation period, could only be explained by the interaction of multipolar components of the surface magnetic field with radial velocity perturbations introduced by radial pulsation. 
 
\ksi~is classified as a $\beta$ Cep pulsator, displaying monoperiodic radial pulsations with a $\sim$5-hour period \citep{Saesen2006}. This pulsational modulation is observed in the X-ray light curve \citep{Oskinova2014} and in the H$\alpha$ emission \citep{Shultz2017}, a phenomenon unique among hot magnetic stars. Furthermore, the star's H$\alpha$ emission exhibits characteristics that suggest it originates within a dynamical magnetosphere \citep{Shultz2017}, a magnetospheric class characteristic of slow rotators. Such a phenomenon is more typically observed within the (more luminous) magnetic O-type star population \citep{Petit2013}, and diverges from all other H$\alpha$-bright magnetic B-type stars, which have rapid rotation and are observed to have optical emission consistent with a centrifugal magnetosphere \citep{Petit2013,Shultz2019d}. \ksi~may therefore be an important transitional object located in the divide between the magnetic O- and B-type stars.

All hot stars show high-ionization ultraviolet (UV) lines caused by outflowing winds, as apparent from their P-Cygni profiles. For non-magnetic stars, these profiles show strong variability predominantly in their absorption component \cite[e.g.][]{Massa1995,Kaper1996,Kaper1999}, which is much more sensitive to changes in the line-of-sight column density than the emission part (which is caused by scattering in all directions). In contrast, the wind-sensitive resonance line profiles of the confirmed magnetic B stars, in particular the He-peculiar stars \citep[e.g.][]{Shore1987,Wade2000,Shultz2019c}, differ significantly in two aspects: the variability extends over the the full P-Cygni profile (including the emission), and the overall shape of the profile is rotationally modulated. Galleries of the UV \cfourfull~line behavior in different types of magnetic B stars have been presented by \citet{Henrichs2001} and \citet{Henrichs2012}. These two properties therefore serve as the strongest indirect indicator of magnetic B stars \citep{Henrichs1993}, and were used to predict the presence of a magnetic field in the B1 IV star $\beta$~Cep \citep[HD 205021;][]{Henrichs2000,Donati2001,Henrichs2013}. A similar prediction from the UV-line behavior was made for the $\beta$~Cep-variable He-strong star V2052 Oph \citep[HD~163472, B2 IV-V;][]{Henrichs1998} after which the field was discovered by \cite{Neiner2003b}. This approach also led to the discovery of the magnetic field of the slowly pulsating B star $\zeta$~Cas \citep[HD 3360, B2 IV;][]{Neiner2003a} and the He-strong star $\sigma$~Lup \citep[HD 127381, B1/B2 V;][]{Henrichs2012}. 

It should be emphasized, however, that the target list to search for magnetic B stars was originally based on the abundance study by \cite{Gies1992}. The authors found a number of near main-sequence B stars (including \ksi, $\beta$~Cep, and $\zeta$~Cas) with an overabundance of nitrogen, which was later confirmed by \cite{Morel2006}. Indeed, the UV \cfour~profile of \ksi~is very anomalous for its spectral type (see Figure~3 of \citealt{Schnerr2008}), which strongly supports the presence of a magnetic field in this star. However, UV rotational modulation has not yet been detected for \ksi, since the only reported UV spectra to date are all consistent with observations of the star near a state of maximum emission \citep{Schnerr2008, Shultz2017}.  

In this paper, we provide an updated spectropolarimetric dataset through the year 2020, and confirm the first definite detection of a negative longitudinal magnetic field in \ksi. We report the results of continued monitoring of the star's unique H$\alpha$ emission, and we present the first new UV spectrum of \ksi~obtained in approximately 40 yrs. In Section \ref{sec:obs}, we describe the observational datasets used in our work. The magnetic analysis is covered in Section \ref{sec:mag}. The H$\alpha$ and UV magnetospheric diagnostics are presented in Section \ref{sec:magdiags}. Finally, we discuss our conclusions in Section \ref{sec:conclusions}.   

\section{Observations}
\label{sec:obs}


\begin{table}
\centering
\caption{Table of Radial Velocity (RV) uncertainties extracted from the full spectropolarimetric sequences and longitudinal magnetic field ($\bz$) measurements obtained during the 2019-2020 epochs. The measurement of the null spectrum ($\nz$) is also reported for each date. Pulsation phases are calculated using the non-linear ephemeris from \citet{Shultz2017}.}
\label{table:bz}
\begin{tabular}{c c c c c c}
\hline
\hline
HJD - & Pulsation & RV & $\bz$ & $\nz$ & Peak \\
2450000 & Phase & (km s$^{-1}$) & (G) &(G) & S/N \\
\hline
8557.76842 & 0.94 & 40.7 & -105 $\pm$ 9 & 0 $\pm$ 9 & 635 \\
8557.77404 & 0.97 & 41.1 & -99 $\pm$ 9 & 5 $\pm$ 9 & 629 \\
8557.83912 & 0.28 & 16.4 & -77 $\pm$ 10 & -8 $\pm$ 10 & 617 \\
8557.84474 & 0.31 & 14.0 & -90 $\pm$ 9 & -15 $\pm$ 9 & 676 \\
8557.85042 & 0.33 & 11.9 & -93 $\pm$ 9 & -9 $\pm$ 9 & 658 \\
8559.76381 & 0.46 & 7.5 & -87 $\pm$ 8 & -18 $\pm$ 8 & 717 \\
8559.76894 & 0.49 & 7.9 & -79 $\pm$ 8 & -5 $\pm$ 8 & 666 \\
8559.78451 & 0.56 & 10.7 & -79 $\pm$ 8 & 7 $\pm$ 8 & 696 \\
8559.78985 & 0.59 & 12.2 & -81 $\pm$ 9 & -11 $\pm$ 9 & 640 \\
8559.79528 & 0.61 & 14.1 & -95 $\pm$ 9 & 6 $\pm$ 9 & 622 \\
8559.81285 & 0.70 & 21.5 & -87 $\pm$ 8 & 7 $\pm$ 8 & 653 \\
8559.82521 & 0.76 & 27.3 & -80 $\pm$ 9 & -2 $\pm$ 9 & 673 \\
8560.71412 & 1.00 & 40.6 & -102 $\pm$ 8 & 8 $\pm$ 8 & 743 \\
8560.72028 & 0.03 & 39.4 & -92 $\pm$ 8 & -1 $\pm$ 8 & 768 \\
8560.72525 & 0.05 & 38.0 & -90 $\pm$ 8 & 2 $\pm$ 8 & 762 \\
8560.76297 & 0.23 & 21.0 & -77 $\pm$ 8 & -1 $\pm$ 8 & 764 \\
8563.74440 & 0.46 & 7.5 & -79 $\pm$ 7 & -10 $\pm$ 7 & 823 \\
8564.84783 & 0.72 & 24.0 & -80 $\pm$ 7 & -6 $\pm$ 7 & 786 \\
8564.85276 & 0.74 & 26.3 & -86 $\pm$ 7 & -1 $\pm$ 7 & 797 \\
9183.05041 & 0.46 & 7.9 & -208 $\pm$ 8 & -6 $\pm$ 8 & 970 \\
9184.00041 & 0.99 & 39.8 & -225 $\pm$ 10 & 5 $\pm$ 10 & 759 \\
9187.02585 & 0.43 & 7.7 & -196 $\pm$ 8 & 4 $\pm$ 7 & 1050 \\
9190.06572 & 0.94 & 40.1 & -210 $\pm$ 8 & 4 $\pm$ 8 & 995 \\
9190.96634 & 0.23 & 19.7 & -213 $\pm$ 8 & -3 $\pm$ 8  & 1017 \\
9191.93719 & 0.86 & 36.9 & -199 $\pm$ 7 & -3 $\pm$ 7  & 1004 \\
9192.02958 & 0.31 & 13.1 & -206 $\pm$ 8 & -1 $\pm$ 8 & 1036 \\
\hline
\hline
\end{tabular}
\end{table}

\subsection{\esp~spectropolarimetry}

Nineteen high-resolution ($R \sim 68,000$) spectropolarimetric sequences were obtained between 2019 March 15 and 2019 March 22 with \esp~at the Canada-France-Hawaii Telescope (CFHT). An additional seven \esp~sequences were obtained between 2020 November 29 and 2020 December 08\footnote{Programme codes 19AC20 and 20BC16.}. The log of these observations is provided in Table \ref{table:bz}. \esp~is an echelle spectropolarimeter that covers a wavelength range of approximately 3700-10500 \AA~across 40 orders. Each observation is comprised of four subexposures, each with a 70 s duration, from which four Stokes $I$ spectra, one Stokes $V$ spectrum, and two null ($N$) spectra are calculated \citep{Donati1997b}. The null spectra are extracted from the four subexposures according to the method outlined in \citet{Donati1997b}, such that any inherent polarization of the source is canceled out. The data were reduced using Upena, the standard pipeline for \esp~data from CFHT. Upena utilizes the Libre-ESpRIT software developed by \citet{Donati1997b}. A detailed description of the reduction and analysis of \esp~data can be found in \citet{Wade2016}. The peak signal-to-noise ($S/N$) ratio of the Stokes $V$ spectra per 1.8~\kms~pixel ranges between 617 and 1050. The mean peak $S/N$ ratio per pixel of the Stokes $I$ spectra is approximately 500 (so the data quality is comparable to previous observing seasons). 

We calculated radial velocity (RV) measurements (Table \ref{table:rv_2020}) from the individual subexposures of the spectropolarimetric sequences, following the multi-line procedure described by \citet{Shultz2017}. The RVs from the 2019 observations were published by \citet{Wade2020}. To calculate pulsation phase, we adopt the non-linear ephemeris ($P_0$ = 0.2095763(1)~d and $\dot{P}$ = 0.0096(5)~s~yr$^{-1}$)~from \citet{Shultz2017}, consistent with the short-term period evolution described by \citet{Wade2020}.   


\subsection{HST ultraviolet spectroscopy}
\label{sec:hstuv}

Forty-two ultraviolet spectra of $\xi^1$ CMa were obtained during a monitoring campaign with the Space Telescope Imaging Spectrograph (STIS) onboard the {\it Hubble Space Telescope} in 2017 February as part of Cycle 24 GO Program 14657 (PI: Oskinova). These data were retrieved from the Mikulski Archive for Space Telescopes (\textsc{MAST}) database\footnote{Access to the \textsc{MAST} database is available at \url{https://archive.stsci.edu/}.}. The observations were broken into 6 visits that occurred over 6 consecutive orbits, with a gap of $\sim$64 minutes between the last exposure of one visit and the start of the first exposure of the next visit. The total time coverage amounted to 8.45 hours, which spanned 1.68 cycles of the $\sim$0.209 d pulsational period of $\xi^1$~CMa. Table~\ref{journal} provides a journal of these observations.

The observing sequence for each visit began with a standard target acquisition to center $\xi^1$ CMa accurately in the 0.3X0.05ND aperture of STIS. The attenuation provided by this aperture was required to mitigate the brightness of the source. Following an ``auto-wavecal'' exposure of the internal Pt-Cr/Ne wavelength calibration lamp, seven consecutive exposures were obtained with the ACCUM mode of the far-ultraviolet MAMA photon-counting detector and the E140M grating. The exposure time for each of the first six spectra was 228 s, but only 218 s for the final spectrum to accommodate orbital constraints. In either case, the degree of ``phase smearing'' of the pulsational period is negligible. The E140M grating provided wavelength coverage from 1140~{\AA} (order 129) to 1709~{\AA} (order 87) with resolving power of $\sim$46,000. The signal-to-noise ratio was $\sim$6 per pixel in the vicinity of 1495~{\AA}.

The spectra were uniformly processed with version 3.4.2 (2018 January 19) of the standard CALSTIS calibration pipeline \citep{stisdatabookv7}, which included steps to: (1) correct for  detector non-linearity; (2) subtract contributions from dark current; (3) apply a flat-field to mitigate pixel-to-pixel sensitivity variations; (4) correct for geometric distortion; and (5) determine and apply corrections to the wavelength zero-point. After a 1D spectrum was extracted for each order, contributions due to scattered light were removed and both the dispersion solution and photometric calibration were applied. As a final step, we stitched together these extracted, calibrated spectra by matching the flux levels in the regions of overlap between successive orders to create a single, merged spectrum for each exposure.

The individual spectra were normalized and coadded into a single spectrum to increase the signal-to-noise ratio. Pulsation-induced variations in the spectra are small compared to the variability due to rotational modulation, therefore a small amount of phase smearing in the coadded spectrum is inconsequential to the purpose of this work (see Section \ref{sec:uvspec} below). 


\subsection{Previously Published Datasets}

We include several previously published datasets in our analysis, described in the sections below.

\subsubsection{\esp ~Spectropolarimetry}

Sixty-four high-resolution spectropolarimetric sequences were obtained with \esp~at CFHT between 2008 January 23 and 2018 July 07 \citep{Silvester2009,Shultz2017, Shultz2018}. A detailed description of the observations from 2008-2017 was given by \citet{Shultz2017}. Observations from the year 2018 were discussed by \citet{Shultz2018}.

\subsubsection{MuSiCoS Spectropolarimetry}

Three high-resolution ($R \sim 35,000$) optical Stokes $V$ spectra were obtained between 2000 February 05 and 2000 February 11 using the MuSiCoS spectropolarimeter, which was mounted on the Bernard Lyot Telescope (TBL) at the Pic du Midi Observatory \citep{Baudrand1992,Donati1999}. Details about the reduction and analysis of these data were given by \citet{Shultz2017}.

\subsubsection{\textsc{CORALIE} Spectroscopy}

A large number of spectra (401) were obtained using the \textsc{CORALIE} optical spectrograph between 2000 February 17 and 2004 October 05. \textsc{CORALIE} is mounted at the Leonhard Euler Telescope at La Silla observatory, and is operated by the Geneva Observatory \citep{Queloz2000,Queloz2001}. These data were first analyzed and published by \citet{Saesen2006}, and were used again by \citet{Shultz2017}.

\subsubsection{IUE Ultraviolet Spectroscopy}

$\xi^1$ CMa was observed 13 times by the Interstellar Ultraviolet Explorer (IUE) in high-resolution ($\lambda/\Delta\lambda \sim 20,000$) mode using the Short Wavelength Prime (SWP) camera. The $S/N$ of all observations is approximately 20. The first observation was acquired in 1978, and the remaining 12 about 142 days later, covering one pulsation cycle with regular phase intervals. The data were reduced with the New Spectral Image Processing System (NEWSIPS; \citealt{Nichols1996}) and were retrieved from the MAST archive\footnote{Available at \url{https://archive.stsci.edu/iue/.}}. We used the absolute calibrated flux, discarded pixels flagged as anomalous, and merged individual spectral orders at the wavelengths where the flux uncertainties of adjacent orders overlapped. The spectra were normalized and coadded into a single spectrum. Since the pulsation amplitude is much smaller than the line widths, pulsation smearing is negligible.

We also utilized seven spectra of $\beta$ Cep obtained with IUE/SWP between 1979 February 24 and 1995 February 14. The data were reduced and post-processed using the same procedure as described above. Since the rotational period of $\beta$ Cep is known (12~d; \citealt{Henrichs1993,Henrichs2013}), we selected the spectra at maximum emission (rotation phase $\phi_{\mathrm{rot}} = 0.0$) and at minimum emission ($\phi_{\mathrm{rot}} = 0.25$) for comparison to the UV spectra of \ksi. For the \nfivefull~spectral line, in order to increase the signal-to-noise ratio, we selected spectra within the range $\phi_{\mathrm{rot}} = 0.0 \pm 0.025$. For each group, the spectra were normalized and coadded into a single spectrum.

\section{Magnetic Analysis}
\label{sec:mag}



\begin{figure*}
\centering
\includegraphics[width=0.896\textwidth]{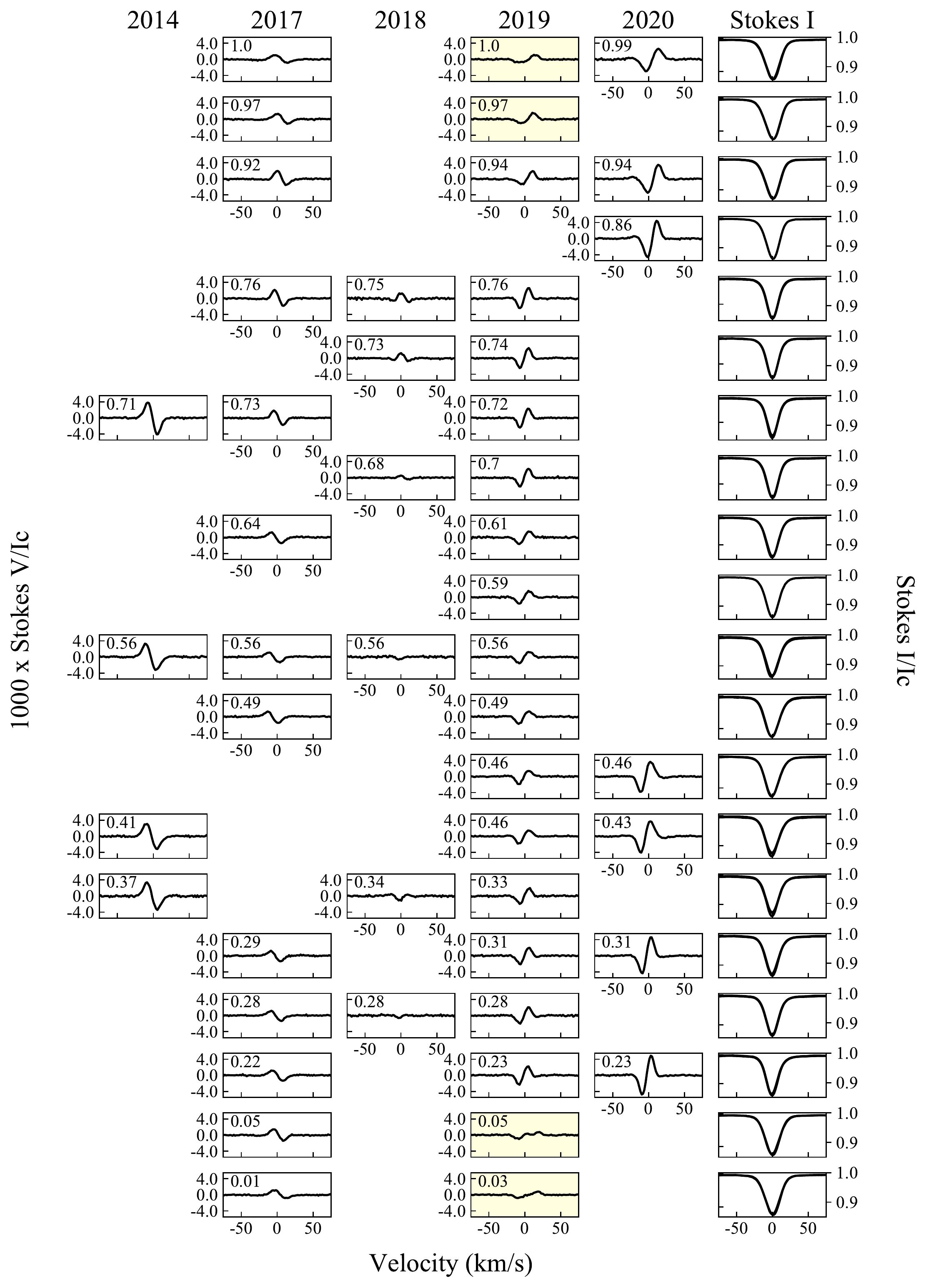}
\caption{Stokes $V$ and Stokes $I$ profiles from several recent observational epochs, shifted to their central velocities, and arranged in order of increasing pulsation phase (indicated in the upper left-hand corner of each subplot). The 2019-2020 Stokes $V$ profiles show negative Zeeman signatures, in contrast to the positive signatures in the 2014-2017 profiles, and the crossover signatures in the 2018 dataset. We highlight (in yellow shading) the 2019 Stokes $V$ profiles near pulsation phase 0.05 that have an unusual shape. The Stokes $I$ profiles obtained at a given pulsation phase are identical between years.} 
\label{fig:lsd}
\end{figure*}
\begin{figure}
\centering
\includegraphics[width=0.47\textwidth]{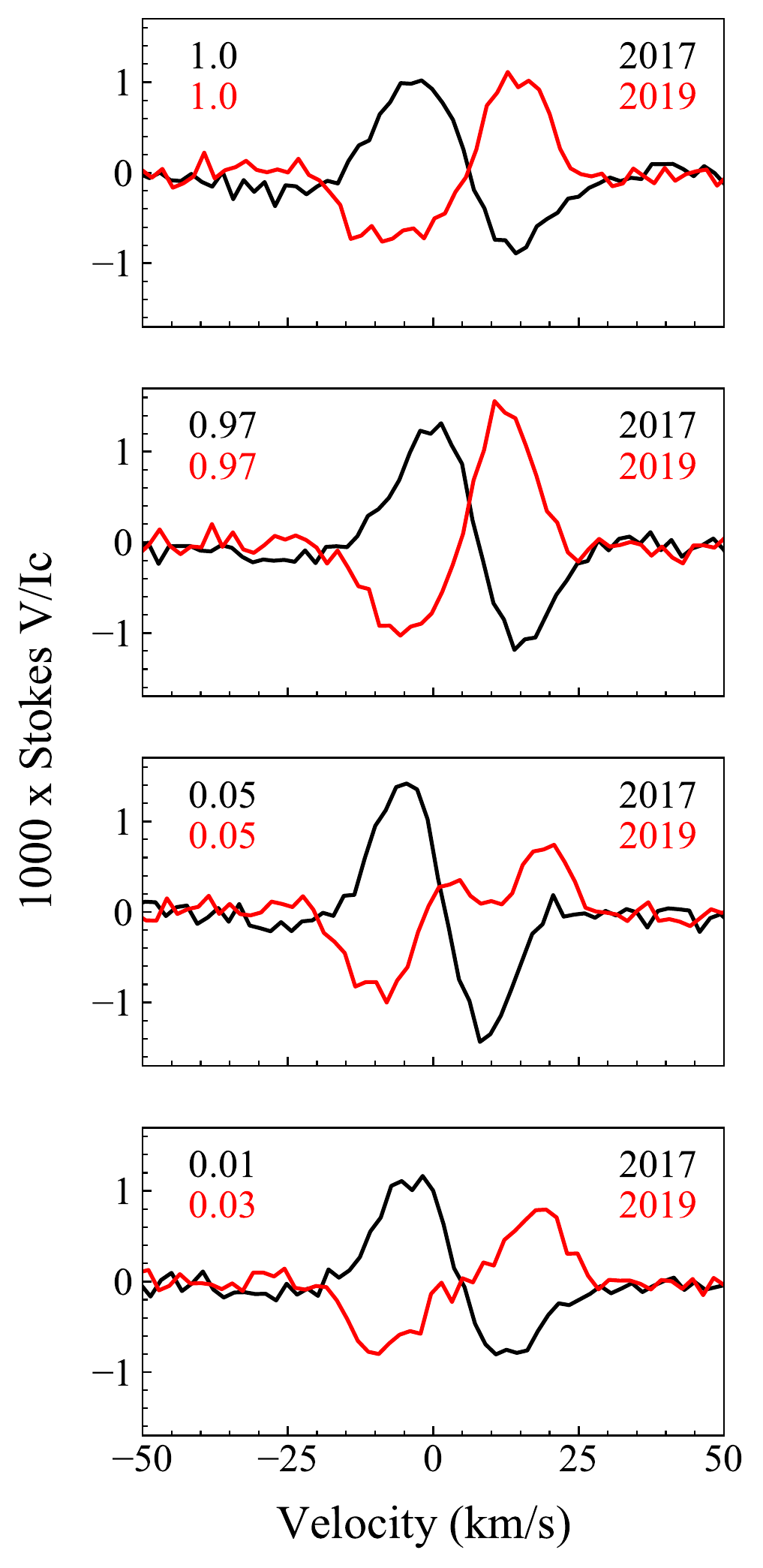}
\caption{A close-up view of the four highlighted panels in Figure~\ref{fig:lsd}. These \stokesv~profiles from 2019 (red) have unusual shapes when compared to the \stokesv profiles from 2017 (black) at similar pulsation phase (indicated in the upper left-hand corner of each subplot).} 
\label{fig:lsd_fourpanel}
\end{figure}

In order to increase the signal-to-noise ratio of the Zeeman signatures present in the \stokesv~profiles of \ksi, we extracted mean line profiles using the Least-Squares Deconvolution (LSD; \citealt{Donati1997b}) method. We applied the same custom Vienna Atomic Line Database (VALD3; \citealt{Piskunov1995, Ryabchikova1997, Ryabchikova2015, Kupka1999, Kupka2000}) line mask as \citet{Shultz2017, Shultz2018}. The LSD profiles are extracted with weights corresponding to a line with $d$=0.1, $g$=1.2, and $\lambda$=500~nm. The individual line profiles were shifted to their respective rest velocities by subtracting the measured radial velocities from Table \ref{table:bz}. We calculated False Alarm Probabilities (FAPs; \citealt{Donati1997b}), which are used to assess detection flag statistics. We used a range of $\pm$30~\kms~around line center. All of the Stokes $V$ profiles yield definite detections (DD; FAP $<10^{-5}$), and all of the null profiles produce non-detections (ND; FAP $>10^{-3}$). This is consistent with previous measurements of $\bz$ and $\nz$; DDs in the null profiles do occur in \ksi, but only when $\bz$ is high \citep{Shultz2017}.

Figure~\ref{fig:lsd} shows the \stokesv~and Stokes $I$ profiles from the 2019 and 2020 spectropolarimetric datasets (new to this paper), compared with profiles from earlier observational datasets (previously published by \citealt{Shultz2017,Shultz2018}). The 2019 observations provide the first unambiguous measurement of a negative longitudinal magnetic field, confirming the inference from the MuSiCoS dataset (formally a non-detection) that both magnetic poles are visible over a rotational cycle. This can also be seen in the reversal of the Stokes $V$ signature in the LSD profiles between the 2017 and 2019 epochs.

The Stokes $V$ profiles show similar morphologies and polarities, and yield similar longitudinal fields, within a given year, consistent with expectations for a long rotation period. The 2019 observations match the amplitude (with reversed polarity) of the corresponding observations in the 2017 dataset that were obtained at the same pulsational phase. Furthermore, the amplitudes of the 2020 profiles are similar to those in the 2014 dataset. Positive Zeeman signatures are seen in the 2014 and 2017 \stokesv~profiles, while negative Zeeman signatures appear in the 2019 and 2020 profiles. Such behavior is consistent with what would be expected in the case of a long rotational period, with the longitudinal field crossing magnetic null in 2018 \citep{Shultz2018}. Any short-term variability of the Stokes $V$ profiles is therefore best explained by pulsation, while the long-term variability is due to rotation.

The 2019 Stokes $V$ profiles near pulsation phase $\phi_{\mathrm{p}} = 0$ have unusual, non-antisymmetric shapes, which are distinct from the observations obtained at pulsation phases just above and below that value. These unusual profiles are shown in Figure~\ref{fig:lsd_fourpanel} (red lines), which provides a close-up view of the highlighted panels in Figure~\ref{fig:lsd}. There is no significant change in the shape of the 2019 Stokes $I$ profiles for pulsation phases near $\phi_{\mathrm{p}} = 0$. Additionally, the unusual shape of the \stokesv~profiles is not seen in the 2017 dataset (Figure~\ref{fig:lsd_fourpanel}, black lines) which also has closely spaced observations in the same range of pulsation phases. 

\citet{Shultz2018} reported the presence of crossover signatures\footnote{The term \textit{crossover signature} is used to describe a Zeeman signature corresponding to the magnetic null. In contrast to the usual antisymmetric Zeeman signature, a crossover signature is symmetric about line center and results from the distribution of Doppler shifts at the stellar surface.} that vary coherently with pulsation phase in the 2018 \stokesv~profiles, which are not anticipated given the negligible $v \sin i$ inferred from \ksi's long rotation period. The authors found that such crossover signatures are the result of a ``radial crossover" effect, arising from the interaction between the velocity field introduced by radial pulsations and the complex topology of the magnetic field. It is therefore possible that the change in the shape of the 2019 Stokes $V$ profile also results from this interaction. Furthermore, because this abnormality is absent from the 2017 \stokesv~profiles, there may be an as-yet unobserved asymmetry in the topology of the field between the north and south magnetic poles. 

The longitudinal magnetic field $\bz$ provides a quantitative measure of the line-of-sight field strength \citep{Mathys1989}. Accordingly, we measured $\bz$ using an integration range of $\pm$ 30 km~s$^{-1}$ around line center, following the procedure from \citet{Shultz2017}, and the LSD scaling mentioned above. The $\bz$ measurement for each polarimetric sequence in the new epoch is reported in Table \ref{table:bz}. All of the $\bz$ values are negative: the error bar-weighted mean of the longitudinal field from the 2019 dataset, $\bz$ = -87 $\pm$ 2 G, decreases to $\bz$ = -207 $\pm$ 3 G in the 2020 dataset. Null profile measurements are all consistent with zero: the weighted mean of the 2019 null profiles is $\nz$ = -3 $\pm$ 2 G. Similarly, the weighted mean of the 2020 null profiles is $\nz$ = -1 $\pm$ 3 G. The mean error in the individual $\bz$ measurements is consistent with the standard deviation of $\bz$ within each observational epoch (about 8 G in all cases).

\begin{figure*}
\centering
\includegraphics[width=\textwidth]{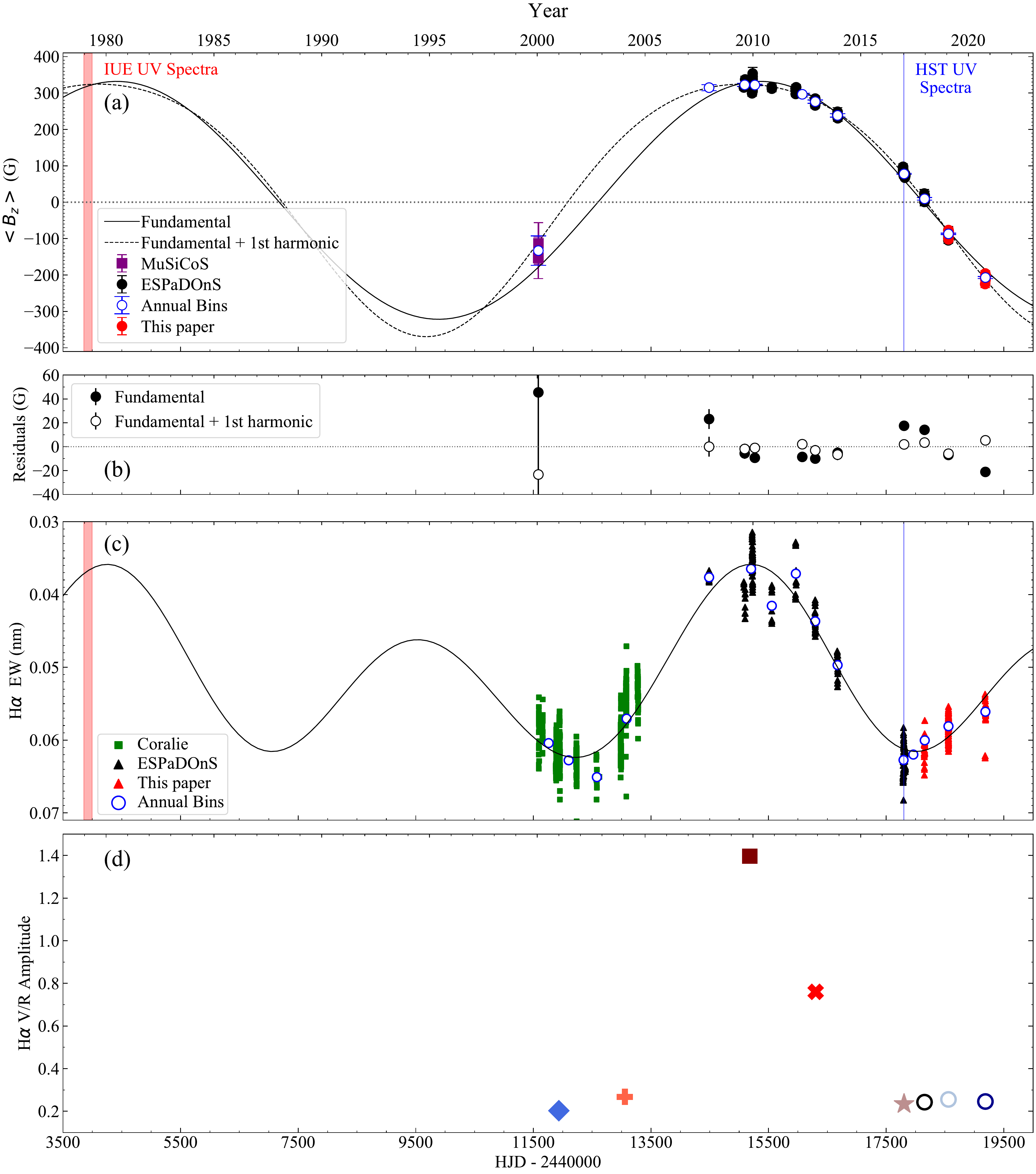}
\caption{\textbf{(a):} $\bz$ as a function of time. Previously-reported observations (purple and black filled circles) are plotted alongside the new 2019-2020 data (red filled circles). The annually-binned weighted means of $\bz$ are indicated by the blue open circles. We also include two fits to the weighted means using a fundamental harmonic model (black solid line) and a first harmonic model (black dashed line). The vertical shading illustrates when the {\it IUE} and {\it HST} UV spectra were obtained. \textbf{(b):} Residuals for the fits to the $\bz$ curve. \textbf{(c):} H$\alpha$ equivalent widths (EWs) as a function of time. The EWs from 2000-2017 were previously published by \citet{Shultz2017}; EWs from 2018-2020 (in red filled triangles) are new to this paper. The annually-binned weighted means of the H$\alpha$ EWs are marked in blue open circles. The black curve shows the best fit second harmonic model. \textbf{(d):} Amplitude of the H$\alpha$ V/R curve as a function of time. The data bins, colors, and symbols match the H$\alpha$ V/R curve shown in Figure~\ref{fig:havr}.} 
\label{fig:bz_fig}
\end{figure*}

Figure~\ref{fig:bz_fig}a shows the $\bz$ curve with a sinusoidal fundamental harmonic fit (black solid line), illustrating what would be expected for a purely dipolar field. From this fit, the sum of the inclination and obliquity angles $i + \beta$ can be constrained \citep{Preston1967}. We calculated Preston's $r$ parameter using the Monte Carlo sampler from \citet{Shultz2019d}, setting the mean $\bz$ variation to $B_0 = 20 \pm 20$ G, and the semi-amplitude of the first harmonic to $B_1 = 1620 \pm 70$ G. We find $ i + \beta = 139^{\circ}$, with 1$\sigma$ of uncertainties of ($+12^{\circ},-15^{\circ}$), which is qualitatively consistent with \citet{Shultz2017}. 
Because \ksi~exhibits departures from a purely dipolar topology, we also produced a first harmonic fit to $\bz$ (i.e. a dipole with a quadrupolar component; black dashed line). Both models assume an ephemeris of HJD$_0$ = 2455220 and a rotation period of 30 yrs, consistent with the best-fit models from \citet{Shultz2017,Shultz2018}.

The first harmonic fit is preferred based on its smaller residuals (Figure~\ref{fig:bz_fig}b). The reduced $\chi^2$ of the fundamental harmonic model is 34, with a p-value of effectively 0, indicating that the deviation from the best fit model is not due to statistical noise. The reduced $\chi^2$ of the first harmonic model is 4.3, indicating a better fit to the data within the statistical possibility due to random noise (although, with a p-value of $10^{-4}$, it is still not statistically compatible with a good fit).

We also calculated the Akaike information criterion score (AIC; \citealt{Akaike1974}) for each model, which considers the number of parameters and the goodness of fit against increased model complexity. The fundamental harmonic model yielded an AIC score of 295, while the first harmonic model produced an AIC score of 83. A lower AIC score indicates the first harmonic fit is the preferred model. 

To evaluate the possibility that the uncertainty on $\bz$ may have been underestimated, we repeated this test, increasing the error on the $\bz$ curve by 50\%. In this case, the first harmonic model is statistically preferred (p-value = 0.075), and the fundamental harmonic model is still statistically excluded (with a p-value of effectively 0). The fundamental harmonic model yielded an AIC score of 162, and the first harmonic model produced an AIC score of 71, indicating that in the unlikely event the error bars were underestimated by 50\% due to unknown sources of systematic error, the conclusion that the first harmonic fit is preferred remains unchanged.

A caveat to this result is that the full rotation cycle of \ksi~has not yet been completely observed. It is likely that the magnetic topology is predominantly dipolar, and the higher order components are small; however, the actual topological configuration cannot be determined at this point. This said, our results are {\em consistent} with the evidence for a complex field, provided by the analyses of the \stokesv~profiles \citep[][and above]{Shultz2018}. Further monitoring of the longitudinal field curve will be required in order to fully constrain the rotational period of \ksi.

\section{Magnetospheric Diagnostics}
\label{sec:magdiags}

The magnetic fields of O- and B-type stars trap the outflowing stellar wind and channel it into a circumstellar magnetosphere that co-rotates with the star. If the star is not rotating fast enough to provide sufficient centrifugal support to the plasma to overcome gravity, the confined wind material will fall back to the stellar surface under the influence of gravity on a dynamical timescale, forming a ``Dynamical Magnetosphere'' \citep[DM;][]{Sundqvist2012,Petit2013}. To date, \ksi~is the only magnetic B-type star observed to have H$\alpha$ emission characteristic of originating within a DM \citep{Shultz2017}. 

Since the magnetic axis is often inclined with respect to the stellar rotation axis, the magnetosphere can be viewed from different angles as the star rotates, leading to variability within the stellar spectrum. Wind-sensitive UV resonance lines (e.g. \sifourfull, \cfourfull, and \nfivefull), as well as the H$\alpha$ line, often bear signatures of this rotational modulation. For example, maximum (minimum) H$\alpha$ emission typically corresponds with the closest approach of the magnetic pole (equator) to an observer's line-of-sight. Such rotational modulation in UV lines has been observed \citep[see e.g.][]{Henrichs2013,Shultz2018b,David-Uraz2021} and modeled \citep[see e.g.][]{Marcolino2013, udDoula2013, Erba2020} in detail for both magnetic O- and B-type stars. 
We present here an analysis of the H$\alpha$ emission of \ksi~and consider it in the context of the observed longitudinal field curve. We also examine the variability of the \sifour, \cfour, and \nfive~doublets in the UV spectra of \ksi, and compare it to the rotational modulation visible in the UV spectra of the similar magnetic pulsator $\beta$ Cep.

\subsection{H\texorpdfstring{$\alpha$}{a} Emission}
\label{sec:halpha}

\begin{figure}
\centering
\includegraphics[width=0.45\textwidth]{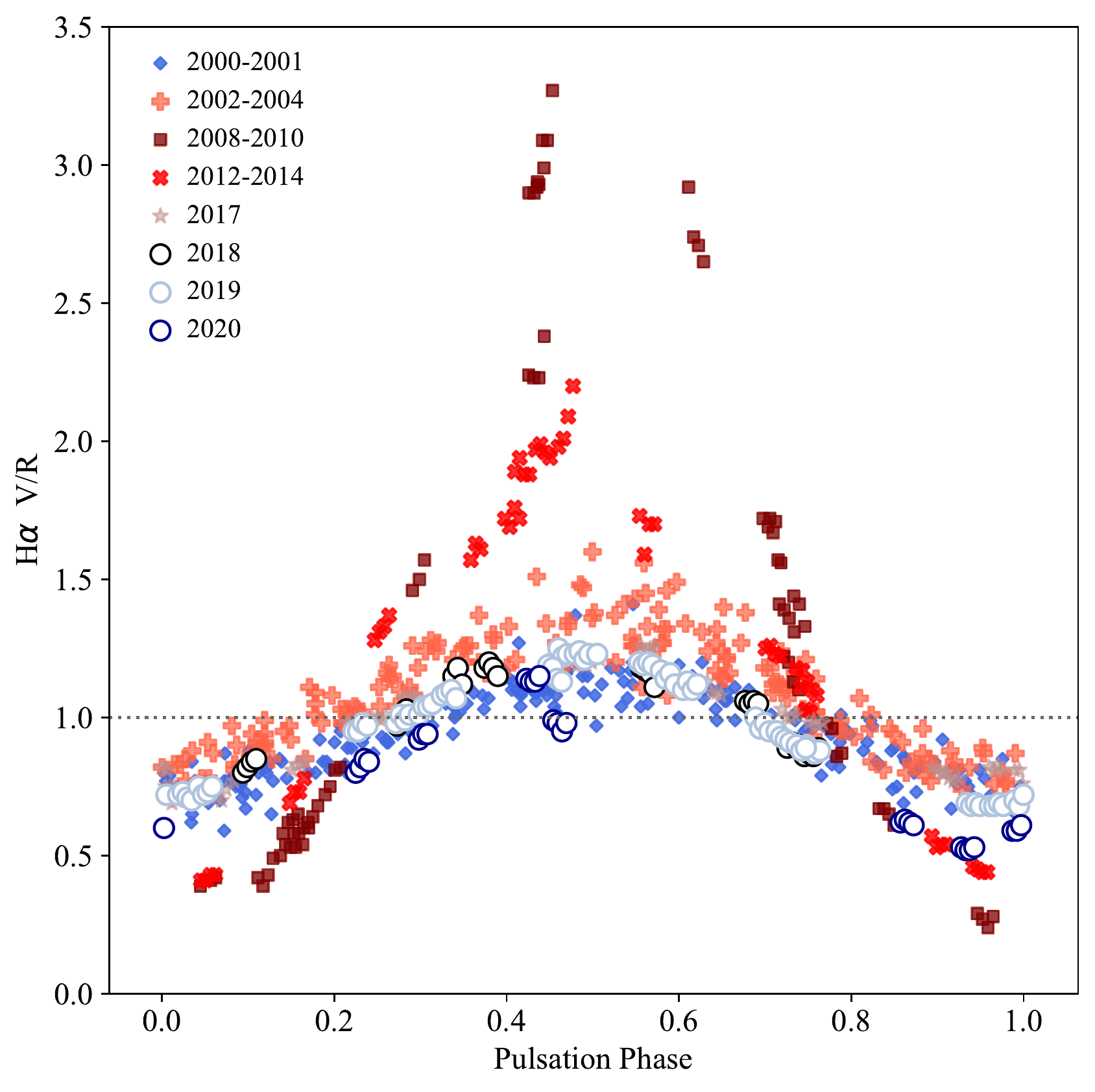}
\caption{H$\alpha$ V/R as a function of pulsation phase. The symbols represent individual observing epochs, and the color indicates proximity to the maximum of the observed longitudinal field curve (dark red is closer to $\langle B_{z,\textrm{max}} \rangle$, dark blue is closer to $\langle B_{z,\textrm{min}} \rangle$).}
\label{fig:havr}
\end{figure}

We calculated the equivalent width (EW) of the H$\alpha$ line in the 2018-2020 observational epochs using an integration range of $\pm 0.4$~nm about line centre. In order to provide a precise measurement of the variation in line strength, we renormalized the line profiles to a consistent continuum by fitting a line to the spectrum within 0.05~nm of the edge of the integration range, then dividing the profile by the line. These measurements are reported in Table \ref{table:appx_halpha}. The EWs from 2000-2017 were previously published by \citet{Shultz2017}.

Figure~\ref{fig:bz_fig}c shows the variation of the H$\alpha$ EW as a function of time. The H$\alpha$ emission is modulated on the pulsation period, which introduces scatter outside the formal error bars in the measurements obtained at any given epoch \citep{Shultz2017}. We fit the annual weighted means (open circles) with a second harmonic sinusoidal model (solid black curve), using the same ephemeris as the fits to the $\bz$ curve. Since both magnetic poles are visible, we expect the EW to follow a double-wave variation \citep[see e.g.][]{Shultz2020}, requiring a minimum of two sinusoidal terms to fit. In practice we found that three terms were necessary to obtain a reasonable fit. While the H$\alpha$ emission provides a well-sampled dataset, it is almost guaranteed to be anharmonic due to its formation within \ksi's magnetosphere. Therefore, a determination of the rotation period from an as-yet incomplete H$\alpha$ timeseries is not strictly reliable.

Overall, the variation of the H$\alpha$ emission shown in Figure~\ref{fig:bz_fig}c is qualitatively consistent with the magnetic data. The minimum of the H$\alpha$ EW (and therefore maximum H$\alpha$ emission) corresponds with the maximum of $\bz$; conversely, the null of $\bz$ corresponds with the maximum H$\alpha$ EW. This double wave variation is expected from a magnetosphere co-rotating with the star in which both poles are viewed. We expect to see an increase in the strength of the H$\alpha$ emission as the negative pole approaches the line of sight.

\citet{Shultz2017} showed that the H$\alpha$ emission profile is highly asymmetric, and that this asymmetry varies coherently with pulsation. They furthermore demonstrated that this could not be explained by the underlying pulsational variation of the photospheric line profile. As a measurement of emission profile asymmetry, we calculated the H$\alpha$ V/R ratio as the ratio of the equivalent width of the blue side of the line to the equivalent width of the red side of the line, using the procedure from \citet{Shultz2017}. We also applied a 3$\sigma$ outlier rejection criterion to all datasets. The V/R ratio provides a quantitative assessment of pulsationally-induced asymmetries in the line profile. 

In Figure~\ref{fig:havr}, we show the H$\alpha$ V/R variability as a function of pulsation phase. The data are separated by observational epoch, and the colors indicate proximity to the maximum of the longitudinal magnetic field (red, blue colors indicating proximity to $\langle B_{z,\textrm{max}} \rangle$, $\langle B_{z,\textrm{min}} \rangle$, respectively). We calculated the amplitudes of the V/R curve for each observational epoch using a least-squares first-order sinusoidal fit to the data; these are shown in Figure~\ref{fig:bz_fig}d. 
The V/R amplitude varies sharply with $\bz$. Figures~\ref{fig:bz_fig}d and \ref{fig:havr} show that the 2008-2010 V/R curve has a considerably larger amplitude than any of the other V/R curves. This occurs at the positive extremum of $\bz$ (Figure~\ref{fig:bz_fig}a). However, from 2010-2013, the V/R curve amplitude sharply decreases while $\bz$ decreases only slightly. The amplitudes of the V/R curve are still comparably small when the longitudinal field is measured near $-200$ G in the \textsc{CORALIE} data (2000) and the newly obtained \esp~data (2020). It appears that the amplitude of the V/R curve only changes significantly near the magnetic poles. We would therefore expect to observe a sharp increase in the V/R curve amplitude (of similar amplitude to the 2008-2010 dataset) during the negative extremum of the longitudinal magnetic field. Continued monitoring will be needed to confirm this behavior.

\subsection{UV Variability}
\label{sec:uvspec}

\begin{figure}
\centering
\includegraphics[width=0.47\textwidth]{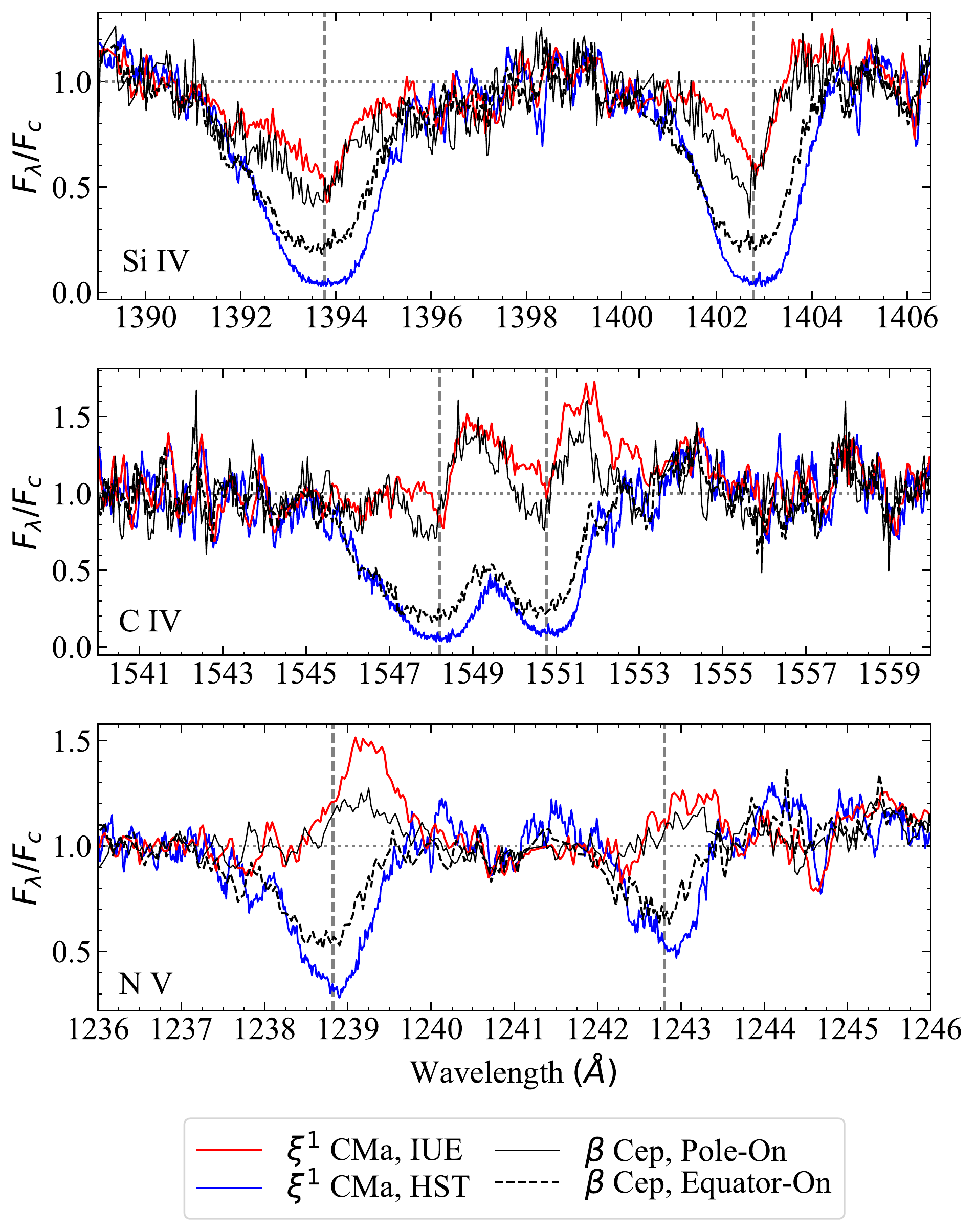}
\caption{{\it IUE} and {\it HST} spectra (black dashed line and solid lines, respectively) of the \sifour, \cfour, and \nfive wind-sensitive UV resonance lines of \ksi, compared with {\it IUE} spectra of the same lines in $\beta$ Cep, obtained at maximum (red lines) and minimum (blue lines) emission.} 
\label{fig:uvspec}
\end{figure}

As noted by \citet{Schnerr2008} and \citet{Shultz2017}, the {\it IUE} UV emission spectra of \ksi~bear a strong resemblance to the archetypal magnetic $\beta$ Cep pulsator $\beta$ Cep (HD 205021, B1 IV) in the \nfive, \sifour, and \cfour~doublets. $\beta$ Cep exhibits a double-wave variation in its UV emission with a rotation period of 12 d \citep{Henrichs1993,Henrichs2013}. In this star, maximum emission occurs at rotation phases $\phi_{\mathrm{rot}} = 0, 0.5$ (corresponding with the extrema of the $\bz$ curve), and minimum emission occurs at rotation phases $\phi_{\mathrm{rot}} = 0.25, 0.75$ (corresponding with the magnetic null).

An initial comparison of these two stars was presented by \citet{Schnerr2008}, who showed the similarity of \ksi's \cfour~doublet to that of $\beta$ Cep, observed at maximum emission. A similar analysis, using the same ({\it IUE}) data as \citet{Schnerr2008}, was given by \citet{Shultz2017} for the \nfivefull~and Al \textsc{iii} $\lambda\lambda$1854, 1863 doublets. The authors came to a similar conclusion, emphasizing the strong resemblance between the {\it IUE} spectrum of \ksi~and $\beta$ Cep's maximum emission state. \citet{Schnerr2008} highlighted that the {\it IUE} spectrum of \ksi~was obtained over a relatively small timescale (approx. 7 h), and so it was unlikely that rotational variability would be observable in \ksi's UV wind lines. \citet{Shultz2017} proposed that this minimal variability resulted from an oblique dipole configuration with a long rotation period, and showed that the UV data for \ksi~were obtained near phase $\phi_{\mathrm{rot}} = 1.0$, assuming their best-fit 30 yr rotation period. This is illustrated in Figures~\ref{fig:bz_fig}a and \ref{fig:bz_fig}c, where we show the time of the {\it IUE} observation (red vertical bar), along with the projected longitudinal field curve assuming our sinusoidal fits with a rotation period of 30 yr.

Since the analysis of the {\it IUE} data by \citet{Schnerr2008} and \citet{Shultz2017}, an additional set of UV spectra was obtained in 2017 using {\it HST} (Figures~\ref{fig:bz_fig}a and \ref{fig:bz_fig}c, blue vertical bar). These data are consistent with the {\it IUE} observations having been obtained near magnetic maximum. In Figure~\ref{fig:uvspec}, we show the \sifour, \cfour, and \nfive~doublets for the {\it IUE} spectrum (red) and new {\it HST} spectrum (blue) of \ksi, in conjunction with the UV spectra of $\beta$ Cep obtained at maximum emission (a pole-on view of the magnetosphere; solid black lines) and minimum emission (an equator-on view of the magnetosphere; dashed black lines). In this figure, the {\it IUE} spectra of \ksi~have been coadded to produce a single, averaged spectrum that ignores smaller line profile variations due to pulsation. The same procedure was used for the {\it HST} spectra of \ksi. Since the pulsational RV amplitude is much smaller than the line width, pulsational phase smearing in a given dataset is negligible. Furthermore, each individual dataset covers at least one pulsational cycle. The dramatic changes between the two datasets therefore cannot be due to pulsation, but must originate from rotation. We also note that in the {\it IUE} spectra, there is no statistically significant variation of the EW with pulsation. Therefore, the amplitude of line strength variation with pulsation is negligible compared with that due to rotation.

The UV spectra of \ksi~exhibit similar line profile morphologies to those of $\beta$ Cep at the high and low states, corresponding to extrema and nulls of the $\bz$ curve, respectively. From the \esp~spectropolarimetric dataset (see Figure~\ref{fig:bz_fig}a), we know the longitudinal field was $\bz \sim 78$ G when the {\it HST} UV data were obtained, which is relatively close to magnetic null. Assuming a 30 yr rotation period for \ksi, the {\it IUE} and {\it HST} spectra were observed approximately 1.25 rotational cycles apart, corresponding with a one-quarter cycle phase difference between the two sets of spectra. The {\it IUE} spectrum would then correspond with the positive extremum of $\bz$ ($\phi_{\mathrm{rot}} \approx 1.0$ as predicted in \citet{Shultz2017}, placing the {\it HST} spectrum near phase $\phi_{\mathrm{rot}} \approx 0.25$). 

The resonance doublets of \ksi~shown in Figure~\ref{fig:uvspec} also exhibit greater absorption (in the {\it HST} spectrum, blue lines) and emission (in the {\it IUE} spectrum, red lines) than those of $\beta$ Cep. This could be due to \ksi's stronger magnetic field (by about a factor of 4; \citealt{Henrichs2013}), or due to its higher effective temperature or larger mass loss rate. Such speculation only accounts for a few of the many factors that influence UV line formation \citep{erba_uvadm}. Further UV observations of \ksi~obtained at different rotational phases will be critical in order to fully characterize the variability of its UV spectral lines.

\section{Conclusions}
\label{sec:conclusions}

We report here the first two unambiguous detections of a negative magnetic field in \ksi~with amplitudes of $\bz = -87 \pm 2$ G and $\bz = -207 \pm 3$ G. This confirms that both of \ksi's magnetic hemispheres are visible over a single rotational cycle. We show that a first harmonic fit to the longitudinal field curve produces better agreement with the data than a pure sinusoidal fit. However, the preferred fit falls short of statistical significance and cannot be taken as a final model, nor should it be interpreted as primary evidence for departures from a purely dipolar topology. Rather, it is consistent with a magnetic topology that is somewhat more complex than a pure dipole, as first suggested in \citet{Shultz2018}. Based on the first harmonic fit depicted in Figure~\ref{fig:bz_fig}a, we predict that the negative extremum of the longitudinal magnetic field will occur in late 2024. Continued, well-sampled spectropolarimetric monitoring will be critical to our ability to provide improved constraints on the rotational period and on the different magnetic topological models.

Even so, \ksi~is still one of the most slowly rotating hot magnetic stars identified to date. A number of A-type stars with similarly long rotation periods have been observed; however, their slow rotation is likely the result of their pre-Main Sequence evolution (as evidenced by their lack of significant winds; see e.g. \citealt{Stepien2002,Mathys2017}). Alternatively, \ksi's extremely slow rotation may be a consequence of the interplay between its main sequence mass-loss and magnetic field\footnote{Recent work \citep[e.g.][]{Shultz2019d, Takahashi2021} has suggested the rotation periods of Ap stars also increase over time (as in hotter stars), which suggests that magnetic braking during the Main Sequence evolutionary phase may also occur among this population.} \citep[e.g.][]{udDoula2009}. \ksi~is therefore a prime target for detailed evolutionary modeling, to determine if there is a way to self consistently understand the evolution of the star, its magnetic field, and its rotation.

The double-wave variation of the H$\alpha$ emission seen in the equivalent width curve is consistent with a view of the circumstellar magnetosphere in which the observer sees both magnetic poles. The emission line asymmetry is observed to vary sharply with $\bz$, therefore a large increase in the amplitude of the V/R ratio should be observable during the negative extremum of the longitudinal magnetic field. Additional observations will also be an important step in ascertaining the long-term behavior of the H$\alpha$ emission.

Variability in the 2019 \stokesv~profiles near pulsation phase $\phi_{\mathrm{p}} = 0$ may arise from the interaction between the star's radial pulsations and complex magnetic field topology. \ksi's purportedly complex field and long rotation period make it an intriguing, but challenging candidate for Zeeman Doppler Imaging (ZDI; \citealt{Donati1997a,Piskunov2002}), which would enable a more in-depth investigation of the interplay between the radial pulsations and the magnetic field, provided a well-sampled spectropolarimetric dataset is available once the rotational period has been fully covered.

It is worth noting that while departures from a purely dipolar topology are detectable in \ksi, this is likely due to the high precision of the available spectropolarimetric data. The large-scale field of \ksi~is still dominated by the dipolar component. This stands in contrast with other magnetic B-type stars such as $\tau$ Sco (HD 149438, B0.2 V) and HD 37776 (B2 V), for which the higher-order moments of the field are dominant \citep[see e.g.][]{Thompson1985,Donati2006b,DonatiLandstreet2009,Kochukhov2011,Kochukhov2015,Kochukhov2016}.   

We present the first new UV spectrum of \ksi~obtained in the last $\sim$40 yr, observed during 2017 when the longitudinal field curve was approaching magnetic null. There is a striking difference between the new UV spectra and those obtained in 1978-1979, which did not exhibit any measurable change over the course of a year. The new UV data show a close similarity to the UV spectrum of $\beta$ Cep obtained near minimum emission. As both magnetic poles are visible, this suggests that \ksi~should show a double-wave variation in its UV and H$\alpha$ emission. This can be confirmed with future monitoring efforts. If the magnetic field of \ksi~is indeed complex, examining rotational modulation in its spectra will be an important and interesting case study in the impact of magnetic fields on early B-type stars.  

\section*{Acknowledgements}

CE gratefully acknowledges support for this work provided by NASA through grant number HST-AR-15794.001-A from the Space Telescope Science Institute, which is operated by AURA, Inc., under NASA contract NAS 5-26555. CE also gratefully acknowledges graduate assistant salary support from the Bartol Research Institute in the Department of Physics and Astronomy at the University of Delaware.

CE and VP gratefully acknowledge support for this work provided by NASA through grant numbers HST-GO-15066, HST-GO-13734, and HST-GO-13629 from the Space Telescope Science Institute, which is operated by AURA, Inc., under NASA contract NAS 5-26555.

VP gratefully acknowledges support from the University of Delaware Research Foundation.

MES gratefully acknowledges support provided by the Annie Jump Cannon Fellowship through the University of Delaware, and endowed by the Mount Cuba Astronomical Observatory.

GAW acknowledges Discovery Grant support from the Natural Sciences and Engineering Research Council (NSERC) of Canada.

Finally, the authors wish to thank the referee, Dr. Colin Folsom, for his insightful comments.

\section*{Data Availability Statement}

This work is based on observations obtained using the \esp~spectropolarimeter at the Canada-France-Hawaii Telescope (CFHT), which is operated by the National Research Council of Canada, the Institut National des Science de l'Univers of the Centre National de la Recherche Scientifique of France, and the University of Hawaii. The data are publically available, and can be accessed via the Canadian Astronomy Data Centre (CADC) at \url{https://www.cadc-ccda.hia-iha.nrc-cnrc.gc.ca}.

This work is also based on observations made with the NASA/ESA {\it Hubble Space Telescope} and the {\it International Ultraviolet Explorer} obtained from the Space Telescope Science Institute, which is operated by the Association of Universities for Research in Astronomy, Inc., under NASA contract NAS 5–26555. These data are available from the Mikulski Archive for Space Telescopes (\textsc{MAST}) database, which can be accessed at \url{https://archive.stsci.edu/}.

Other observational datasets included in this work were obtained using the MuSiCoS spectropolarimeter on TBL at Pic du Midi Observatory and the \textsc{CORALIE} spectrograph on the Leonhard Euler Telescope at La Silla Observatory, ESO Chile. These data can be provided by authors upon request.

This work makes use of the VALD database, operated at Uppsala University, the Institute of Astronomy RAS in Moscow, and the University of Vienna. The VALD database can be accessed via the website at \url{http://vald.astro.uu.se/}.


\bibliographystyle{mnras}
\bibliography{database_Erba} 

\begin{thebibliography}{}
\makeatletter
\relax
\def\mn@urlcharsother{\let\do\@makeother \do\$\do\&\do\#\do\^\do\_\do\%\do\~}
\def\mn@doi{\begingroup\mn@urlcharsother \@ifnextchar [ {\mn@doi@}
  {\mn@doi@[]}}
\def\mn@doi@[#1]#2{\def\@tempa{#1}\ifx\@tempa\@empty \href
  {http://dx.doi.org/#2} {doi:#2}\else \href {http://dx.doi.org/#2} {#1}\fi
  \endgroup}
\def\mn@eprint#1#2{\mn@eprint@#1:#2::\@nil}
\def\mn@eprint@arXiv#1{\href {http://arxiv.org/abs/#1} {{\tt arXiv:#1}}}
\def\mn@eprint@dblp#1{\href {http://dblp.uni-trier.de/rec/bibtex/#1.xml}
  {dblp:#1}}
\def\mn@eprint@#1:#2:#3:#4\@nil{\def\@tempa {#1}\def\@tempb {#2}\def\@tempc
  {#3}\ifx \@tempc \@empty \let \@tempc \@tempb \let \@tempb \@tempa \fi \ifx
  \@tempb \@empty \def\@tempb {arXiv}\fi \@ifundefined
  {mn@eprint@\@tempb}{\@tempb:\@tempc}{\expandafter \expandafter \csname
  mn@eprint@\@tempb\endcsname \expandafter{\@tempc}}}

\bibitem[\protect\citeauthoryear{{Akaike}}{{Akaike}}{1974}]{Akaike1974}
{Akaike} H.,  1974, IEEE Transactions on Automatic Control, \href
  {https://ui.adsabs.harvard.edu/abs/1974ITAC...19..716A} {19, 716}

\bibitem[\protect\citeauthoryear{{Baudrand} \& {Bohm}}{{Baudrand} \&
  {Bohm}}{1992}]{Baudrand1992}
{Baudrand} J.,  {Bohm} T.,  1992, \aap, \href
  {https://ui.adsabs.harvard.edu/abs/1992A&A...259..711B} {259, 711}

\bibitem[\protect\citeauthoryear{{David-Uraz}, {Petit}, {Shultz}, {Fullerton},
  {Erba}, {Keszthelyi}, {Seadrow}  \& {Wade}}{{David-Uraz}
  et~al.}{2021}]{David-Uraz2021}
{David-Uraz} A.,  {Petit} V.,  {Shultz} M.~E.,  {Fullerton} A.~W.,  {Erba} C.,
  {Keszthelyi} Z.,  {Seadrow} S.,   {Wade} G.~A.,  2021, \mn@doi [\mnras]
  {10.1093/mnras/staa3768}, \href
  {https://ui.adsabs.harvard.edu/abs/2021MNRAS.501.2677D} {501, 2677}

\bibitem[\protect\citeauthoryear{{Donati} \& {Brown}}{{Donati} \&
  {Brown}}{1997}]{Donati1997a}
{Donati} J.~F.,  {Brown} S.~F.,  1997, \aap, \href
  {https://ui.adsabs.harvard.edu/abs/1997A&A...326.1135D} {326, 1135}

\bibitem[\protect\citeauthoryear{{Donati} \& {Landstreet}}{{Donati} \&
  {Landstreet}}{2009}]{DonatiLandstreet2009}
{Donati} J.~F.,  {Landstreet} J.~D.,  2009, \mn@doi [\araa]
  {10.1146/annurev-astro-082708-101833}, \href
  {https://ui.adsabs.harvard.edu/abs/2009ARA&A..47..333D} {47, 333}

\bibitem[\protect\citeauthoryear{{Donati}, {Semel}, {Carter}, {Rees}  \&
  {Collier Cameron}}{{Donati} et~al.}{1997}]{Donati1997b}
{Donati} J.~F.,  {Semel} M.,  {Carter} B.~D.,  {Rees} D.~E.,   {Collier
  Cameron} A.,  1997, \mn@doi [\mnras] {10.1093/mnras/291.4.658}, \href
  {https://ui.adsabs.harvard.edu/abs/1997MNRAS.291..658D} {291, 658}

\bibitem[\protect\citeauthoryear{{Donati}, {Catala}, {Wade}, {Gallou},
  {Delaigue}  \& {Rabou}}{{Donati} et~al.}{1999}]{Donati1999}
{Donati} J.~F.,  {Catala} C.,  {Wade} G.~A.,  {Gallou} G.,  {Delaigue} G.,
  {Rabou} P.,  1999, \mn@doi [\aaps] {10.1051/aas:1999130}, \href
  {https://ui.adsabs.harvard.edu/abs/1999A&AS..134..149D} {134, 149}

\bibitem[\protect\citeauthoryear{{Donati}, {Wade}, {Babel}, {Henrichs}, {de
  Jong}  \& {Harries}}{{Donati} et~al.}{2001}]{Donati2001}
{Donati} J.~F.,  {Wade} G.~A.,  {Babel} J.,  {Henrichs} H.~f.,  {de Jong}
  J.~A.,   {Harries} T.~J.,  2001, \mn@doi [\mnras]
  {10.1111/j.1365-2966.2001.04713.x}, \href
  {https://ui.adsabs.harvard.edu/abs/2001MNRAS.326.1265D} {326, 1265}

\bibitem[\protect\citeauthoryear{{Donati} et~al.,}{{Donati}
  et~al.}{2006}]{Donati2006b}
{Donati} J.~F.,  et~al., 2006, \mn@doi [\mnras]
  {10.1111/j.1365-2966.2006.10558.x}, \href
  {https://ui.adsabs.harvard.edu/abs/2006MNRAS.370..629D} {370, 629}

\bibitem[\protect\citeauthoryear{{Erba}, {Petit}, {David-Uraz}  \&
  {Fullerton}}{{Erba} et~al.}{2020}]{Erba2020}
{Erba} C.,  {Petit} V.,  {David-Uraz} A.,   {Fullerton} A.,  2020, in {Wade}
  G.,  {Alecian} E.,  {Bohlender} D.,   {Sigut} A.,  eds, ~ Vol. 11, Stellar
  Magnetism: A Workshop in Honour of the Career and Contributions of John D.
  Landstreet. pp 74--80 (\mn@eprint {arXiv} {1912.08748})

\bibitem[\protect\citeauthoryear{{Erba} et~al.,}{{Erba}
  et~al.}{2021}]{erba_uvadm}
{Erba} C.,  et~al., 2021, \mnras~(submitted)

\bibitem[\protect\citeauthoryear{{Gies} \& {Lambert}}{{Gies} \&
  {Lambert}}{1992}]{Gies1992}
{Gies} D.~R.,  {Lambert} D.~L.,  1992, \mn@doi [\apj] {10.1086/171116}, \href
  {https://ui.adsabs.harvard.edu/abs/1992ApJ...387..673G} {387, 673}

\bibitem[\protect\citeauthoryear{{Grunhut} et~al.,}{{Grunhut}
  et~al.}{2017}]{Grunhut2017}
{Grunhut} J.~H.,  et~al., 2017, \mn@doi [\mnras] {10.1093/mnras/stw2743}, \href
  {https://ui.adsabs.harvard.edu/abs/2017MNRAS.465.2432G} {465, 2432}

\bibitem[\protect\citeauthoryear{{Henrichs}}{{Henrichs}}{2001}]{Henrichs2001}
{Henrichs} H.~F.,  2001, in {Mathys} G.,  {Solanki} S.~K.,   {Wickramasinghe}
  D.~T.,  eds,  Astronomical Society of the Pacific Conference Series Vol. 248,
  Magnetic Fields Across the Hertzsprung-Russell Diagram. p.~393

\bibitem[\protect\citeauthoryear{{Henrichs} et~al.}{{Henrichs}
  et~al.}{1993}]{Henrichs1993}
{Henrichs} H.~F.,  et~al., 1993, in {Nemec} J.~M.,  {Matthews} J.~M.,  eds, IAU
  Colloq. 139: New Perspectives on Stellar Pulsation and Pulsating Variable
  Stars. p.~186

\bibitem[\protect\citeauthoryear{{Henrichs} et~al.,}{{Henrichs}
  et~al.}{1998}]{Henrichs1998}
{Henrichs} H.~F.,  et~al., 1998, in {Wamsteker} W.,  {Gonzalez Riestra} R.,
  {Harris} B.,  eds,  ESA Special Publication Vol. 413, Ultraviolet
  Astrophysics Beyond the IUE Final Archive. p.~157

\bibitem[\protect\citeauthoryear{{Henrichs} et~al.,}{{Henrichs}
  et~al.}{2000}]{Henrichs2000}
{Henrichs} H.~F.,  et~al., 2000, in {Smith} M.~A.,  {Henrichs} H.~F.,
  {Fabregat} J.,  eds,  Astronomical Society of the Pacific Conference Series
  Vol. 214, IAU Colloq. 175: The Be Phenomenon in Early-Type Stars. p.~324

\bibitem[\protect\citeauthoryear{{Henrichs} et~al.,}{{Henrichs}
  et~al.}{2012}]{Henrichs2012}
{Henrichs} H.~F.,  et~al., 2012, \mn@doi [\aap] {10.1051/0004-6361/201219632},
  \href {http://adsabs.harvard.edu/abs/2012A%26A...545A.119H} {545, A119}

\bibitem[\protect\citeauthoryear{{Henrichs} et~al.,}{{Henrichs}
  et~al.}{2013}]{Henrichs2013}
{Henrichs} H.~F.,  et~al., 2013, \mn@doi [\aap] {10.1051/0004-6361/201321584},
  \href {https://ui.adsabs.harvard.edu/abs/2013A&A...555A..46H} {555, A46}

\bibitem[\protect\citeauthoryear{{Hubrig}, {Briquet}, {Sch{\"o}ller}, {De Cat},
  {Mathys}  \& {Aerts}}{{Hubrig} et~al.}{2006}]{Hubrig2006}
{Hubrig} S.,  {Briquet} M.,  {Sch{\"o}ller} M.,  {De Cat} P.,  {Mathys} G.,
  {Aerts} C.,  2006, \mn@doi [\mnras] {10.1111/j.1745-3933.2006.00175.x}, \href
  {https://ui.adsabs.harvard.edu/abs/2006MNRAS.369L..61H} {369, L61}

\bibitem[\protect\citeauthoryear{{Hubrig}, {Ilyin}, {Sch{\"o}ller}, {Briquet},
  {Morel}  \& {De Cat}}{{Hubrig} et~al.}{2011}]{Hubrig2011}
{Hubrig} S.,  {Ilyin} I.,  {Sch{\"o}ller} M.,  {Briquet} M.,  {Morel} T.,   {De
  Cat} P.,  2011, \mn@doi [\apjl] {10.1088/2041-8205/726/1/L5}, \href
  {https://ui.adsabs.harvard.edu/abs/2011ApJ...726L...5H} {726, L5}

\bibitem[\protect\citeauthoryear{{J{\"a}rvinen}, {Hubrig}, {Sch{\"o}ller}  \&
  {Ilyin}}{{J{\"a}rvinen} et~al.}{2018}]{Jarvinen2018}
{J{\"a}rvinen} S.~P.,  {Hubrig} S.,  {Sch{\"o}ller} M.,   {Ilyin} I.,  2018,
  \mn@doi [\na] {10.1016/j.newast.2018.01.005}, \href
  {https://ui.adsabs.harvard.edu/abs/2018NewA...62...37J} {62, 37}

\bibitem[\protect\citeauthoryear{{Kaper}, {Henrichs}, {Nichols}, {Snoek},
  {Volten}  \& {Zwarthoed}}{{Kaper} et~al.}{1996}]{Kaper1996}
{Kaper} L.,  {Henrichs} H.~F.,  {Nichols} J.~S.,  {Snoek} L.~C.,  {Volten} H.,
   {Zwarthoed} G.~A.~A.,  1996, \aaps, 116, 257

\bibitem[\protect\citeauthoryear{{Kaper}, {Henrichs}, {Nichols}  \&
  {Telting}}{{Kaper} et~al.}{1999}]{Kaper1999}
{Kaper} L.,  {Henrichs} H.~F.,  {Nichols} J.~S.,   {Telting} J.~H.,  1999,
  \aap, 344, 231

\bibitem[\protect\citeauthoryear{{Keszthelyi}, {Meynet}, {Georgy}, {Wade},
  {Petit}  \& {David-Uraz}}{{Keszthelyi} et~al.}{2019}]{Keszthelyi2019}
{Keszthelyi} Z.,  {Meynet} G.,  {Georgy} C.,  {Wade} G.~A.,  {Petit} V.,
  {David-Uraz} A.,  2019, \mn@doi [\mnras] {10.1093/mnras/stz772}, \href
  {https://ui.adsabs.harvard.edu/abs/2019MNRAS.485.5843K} {485, 5843}

\bibitem[\protect\citeauthoryear{{Kochukhov}}{{Kochukhov}}{2015}]{Kochukhov2015}
{Kochukhov} O.,  2015, \mn@doi [\aap] {10.1051/0004-6361/201526318}, \href
  {https://ui.adsabs.harvard.edu/abs/2015A&A...580A..39K} {580, A39}

\bibitem[\protect\citeauthoryear{{Kochukhov} \& {Wade}}{{Kochukhov} \&
  {Wade}}{2016}]{Kochukhov2016}
{Kochukhov} O.,  {Wade} G.~A.,  2016, \mn@doi [\aap]
  {10.1051/0004-6361/201527454}, \href
  {https://ui.adsabs.harvard.edu/abs/2016A&A...586A..30K} {586, A30}

\bibitem[\protect\citeauthoryear{{Kochukhov}, {Lundin}, {Romanyuk}  \&
  {Kudryavtsev}}{{Kochukhov} et~al.}{2011}]{Kochukhov2011}
{Kochukhov} O.,  {Lundin} A.,  {Romanyuk} I.,   {Kudryavtsev} D.,  2011,
  \mn@doi [\apj] {10.1088/0004-637X/726/1/24}, \href
  {https://ui.adsabs.harvard.edu/abs/2011ApJ...726...24K} {726, 24}

\bibitem[\protect\citeauthoryear{{Kupka}, {Piskunov}, {Ryabchikova}, {Stempels}
   \& {Weiss}}{{Kupka} et~al.}{1999}]{Kupka1999}
{Kupka} F.,  {Piskunov} N.,  {Ryabchikova} T.~A.,  {Stempels} H.~C.,   {Weiss}
  W.~W.,  1999, \mn@doi [\aaps] {10.1051/aas:1999267}, \href
  {https://ui.adsabs.harvard.edu/abs/1999A&AS..138..119K} {138, 119}

\bibitem[\protect\citeauthoryear{{Kupka}, {Ryabchikova}, {Piskunov}, {Stempels}
   \& {Weiss}}{{Kupka} et~al.}{2000}]{Kupka2000}
{Kupka} F.~G.,  {Ryabchikova} T.~A.,  {Piskunov} N.~E.,  {Stempels} H.~C.,
  {Weiss} W.~W.,  2000, \mn@doi [Baltic Astronomy] {10.1515/astro-2000-0420},
  \href {https://ui.adsabs.harvard.edu/abs/2000BaltA...9..590K} {9, 590}

\bibitem[\protect\citeauthoryear{{Marcolino}, {Bouret}, {Sundqvist}, {Walborn},
  {Fullerton}, {Howarth}, {Wade}  \& {ud-Doula}}{{Marcolino}
  et~al.}{2013}]{Marcolino2013}
{Marcolino} W.~L.~F.,  {Bouret} J.~C.,  {Sundqvist} J.~O.,  {Walborn} N.~R.,
  {Fullerton} A.~W.,  {Howarth} I.~D.,  {Wade} G.~A.,   {ud-Doula} A.,  2013,
  \mn@doi [\mnras] {10.1093/mnras/stt323}, \href
  {https://ui.adsabs.harvard.edu/abs/2013MNRAS.431.2253M} {431, 2253}

\bibitem[\protect\citeauthoryear{{Massa} et~al.,}{{Massa}
  et~al.}{1995}]{Massa1995}
{Massa} D.,  et~al., 1995, \mn@doi [\apjl] {10.1086/309707}, \href
  {http://adsabs.harvard.edu/abs/1995ApJ...452L..53M} {452, L53}

\bibitem[\protect\citeauthoryear{{Mathys}}{{Mathys}}{1989}]{Mathys1989}
{Mathys} G.,  1989, \fcp, \href
  {https://ui.adsabs.harvard.edu/abs/1989FCPh...13..143M} {13, 143}

\bibitem[\protect\citeauthoryear{{Mathys}}{{Mathys}}{2017}]{Mathys2017}
{Mathys} G.,  2017, \mn@doi [\aap] {10.1051/0004-6361/201628429}, \href
  {https://ui.adsabs.harvard.edu/abs/2017A&A...601A..14M} {601, A14}

\bibitem[\protect\citeauthoryear{{Morel}, {Butler}, {Aerts}, {Neiner}  \&
  {Briquet}}{{Morel} et~al.}{2006}]{Morel2006}
{Morel} T.,  {Butler} K.,  {Aerts} C.,  {Neiner} C.,   {Briquet} M.,  2006,
  \mn@doi [\aap] {10.1051/0004-6361:20065171}, \href
  {https://ui.adsabs.harvard.edu/abs/2006A&A...457..651M} {457, 651}

\bibitem[\protect\citeauthoryear{{Morel} et~al.,}{{Morel}
  et~al.}{2015}]{Morel2015}
{Morel} T.,  et~al., 2015, in {Meynet} G.,  {Georgy} C.,  {Groh} J.,   {Stee}
  P.,  eds,  IAU Symposium Vol. 307, New Windows on Massive Stars. pp 342--347
  (\mn@eprint {arXiv} {1408.2100}), \mn@doi{10.1017/S1743921314007054}

\bibitem[\protect\citeauthoryear{{Naz{\'e}}, {Ud-Doula}, {Spano}, {Rauw}, {De
  Becker}  \& {Walborn}}{{Naz{\'e}} et~al.}{2010}]{Naze2010}
{Naz{\'e}} Y.,  {Ud-Doula} A.,  {Spano} M.,  {Rauw} G.,  {De Becker} M.,
  {Walborn} N.~R.,  2010, \mn@doi [\aap] {10.1051/0004-6361/201014333}, \href
  {https://ui.adsabs.harvard.edu/abs/2010A&A...520A..59N} {520, A59}

\bibitem[\protect\citeauthoryear{{Neiner}, {Geers}, {Henrichs}, {Floquet},
  {Fr{\'e}mat}, {Hubert}, {Preuss}  \& {Wiersema}}{{Neiner}
  et~al.}{2003a}]{Neiner2003a}
{Neiner} C.,  {Geers} V.~C.,  {Henrichs} H.~F.,  {Floquet} M.,  {Fr{\'e}mat}
  Y.,  {Hubert} A.-M.,  {Preuss} O.,   {Wiersema} K.,  2003a, \mn@doi [\aap]
  {10.1051/0004-6361:20030742}, \href
  {http://adsabs.harvard.edu/abs/2003A%26A...406.1019N} {406, 1019}

\bibitem[\protect\citeauthoryear{{Neiner} et~al.,}{{Neiner}
  et~al.}{2003b}]{Neiner2003b}
{Neiner} C.,  et~al., 2003b, \mn@doi [\aap] {10.1051/0004-6361:20031342}, \href
  {http://adsabs.harvard.edu/abs/2003A%26A...411..565N} {411, 565}

\bibitem[\protect\citeauthoryear{{Nichols} \& {Linsky}}{{Nichols} \&
  {Linsky}}{1996}]{Nichols1996}
{Nichols} J.~S.,  {Linsky} J.~L.,  1996, \mn@doi [\aj] {10.1086/117803}, \href
  {https://ui.adsabs.harvard.edu/abs/1996AJ....111..517N} {111, 517}

\bibitem[\protect\citeauthoryear{{Oskinova}, {Naz{\'e}}, {Todt},
  {Huenemoerder}, {Ignace}, {Hubrig}  \& {Hamann}}{{Oskinova}
  et~al.}{2014}]{Oskinova2014}
{Oskinova} L.~M.,  {Naz{\'e}} Y.,  {Todt} H.,  {Huenemoerder} D.~P.,  {Ignace}
  R.,  {Hubrig} S.,   {Hamann} W.-R.,  2014, \mn@doi [Nature Communications]
  {10.1038/ncomms5024}, \href
  {https://ui.adsabs.harvard.edu/abs/2014NatCo...5.4024O} {5, 4024}

\bibitem[\protect\citeauthoryear{{Petit} et~al.,}{{Petit}
  et~al.}{2013}]{Petit2013}
{Petit} V.,  et~al., 2013, \mn@doi [\mnras] {10.1093/mnras/sts344}, \href
  {https://ui.adsabs.harvard.edu/abs/2013MNRAS.429..398P} {429, 398}

\bibitem[\protect\citeauthoryear{{Petit} et~al.,}{{Petit}
  et~al.}{2017}]{Petit2017}
{Petit} V.,  et~al., 2017, \mn@doi [\mnras] {10.1093/mnras/stw3126}, \href
  {https://ui.adsabs.harvard.edu/abs/2017MNRAS.466.1052P} {466, 1052}

\bibitem[\protect\citeauthoryear{{Petit} et~al.,}{{Petit}
  et~al.}{2019}]{Petit2019}
{Petit} V.,  et~al., 2019, \mn@doi [\mnras] {10.1093/mnras/stz2469}, \href
  {https://ui.adsabs.harvard.edu/abs/2019MNRAS.489.5669P} {489, 5669}

\bibitem[\protect\citeauthoryear{{Piskunov} \& {Kochukhov}}{{Piskunov} \&
  {Kochukhov}}{2002}]{Piskunov2002}
{Piskunov} N.,  {Kochukhov} O.,  2002, \mn@doi [\aap]
  {10.1051/0004-6361:20011517}, \href
  {https://ui.adsabs.harvard.edu/abs/2002A&A...381..736P} {381, 736}

\bibitem[\protect\citeauthoryear{{Piskunov}, {Kupka}, {Ryabchikova}, {Weiss}
  \& {Jeffery}}{{Piskunov} et~al.}{1995}]{Piskunov1995}
{Piskunov} N.~E.,  {Kupka} F.,  {Ryabchikova} T.~A.,  {Weiss} W.~W.,
  {Jeffery} C.~S.,  1995, \aaps, \href
  {https://ui.adsabs.harvard.edu/abs/1995A&AS..112..525P} {112, 525}

\bibitem[\protect\citeauthoryear{{Preston}}{{Preston}}{1967}]{Preston1967}
{Preston} G.~W.,  1967, \mn@doi [\apj] {10.1086/149358}, \href
  {https://ui.adsabs.harvard.edu/abs/1967ApJ...150..547P} {150, 547}

\bibitem[\protect\citeauthoryear{{Queloz} et~al.,}{{Queloz}
  et~al.}{2000}]{Queloz2000}
{Queloz} D.,  et~al., 2000, \aap, \href
  {https://ui.adsabs.harvard.edu/abs/2000A&A...354...99Q} {354, 99}

\bibitem[\protect\citeauthoryear{{Queloz} et~al.,}{{Queloz}
  et~al.}{2001}]{Queloz2001}
{Queloz} D.,  et~al., 2001, The Messenger, \href
  {https://ui.adsabs.harvard.edu/abs/2001Msngr.105....1Q} {105, 1}

\bibitem[\protect\citeauthoryear{{Ryabchikova}, {Piskunov}, {Kupka}  \&
  {Weiss}}{{Ryabchikova} et~al.}{1997}]{Ryabchikova1997}
{Ryabchikova} T.~A.,  {Piskunov} N.~E.,  {Kupka} F.,   {Weiss} W.~W.,  1997,
  \mn@doi [Baltic Astronomy] {10.1515/astro-1997-0216}, \href
  {https://ui.adsabs.harvard.edu/abs/1997BaltA...6..244R} {6, 244}

\bibitem[\protect\citeauthoryear{{Ryabchikova}, {Piskunov}, {Kurucz},
  {Stempels}, {Heiter}, {Pakhomov}  \& {Barklem}}{{Ryabchikova}
  et~al.}{2015}]{Ryabchikova2015}
{Ryabchikova} T.,  {Piskunov} N.,  {Kurucz} R.~L.,  {Stempels} H.~C.,  {Heiter}
  U.,  {Pakhomov} Y.,   {Barklem} P.~S.,  2015, \mn@doi [\physscr]
  {10.1088/0031-8949/90/5/054005}, \href
  {https://ui.adsabs.harvard.edu/abs/2015PhyS...90e4005R} {90, 054005}

\bibitem[\protect\citeauthoryear{{Saesen}, {Briquet}  \& {Aerts}}{{Saesen}
  et~al.}{2006}]{Saesen2006}
{Saesen} S.,  {Briquet} M.,   {Aerts} C.,  2006, \mn@doi [Communications in
  Asteroseismology] {10.1553/cia147s109}, \href
  {https://ui.adsabs.harvard.edu/abs/2006CoAst.147..109S} {147, 109}

\bibitem[\protect\citeauthoryear{{Schnerr} et~al.,}{{Schnerr}
  et~al.}{2008}]{Schnerr2008}
{Schnerr} R.~S.,  et~al., 2008, \mn@doi [\aap] {10.1051/0004-6361:20077740},
  \href {https://ui.adsabs.harvard.edu/abs/2008A&A...483..857S} {483, 857}

\bibitem[\protect\citeauthoryear{{Shore}}{{Shore}}{1987}]{Shore1987}
{Shore} S.~N.,  1987, \mn@doi [\aj] {10.1086/114511}, \href
  {https://ui.adsabs.harvard.edu/abs/1987AJ.....94..731S} {94, 731}

\bibitem[\protect\citeauthoryear{{Shultz} \& {Wade}}{{Shultz} \&
  {Wade}}{2017}]{Shultz2017a}
{Shultz} M.,  {Wade} G.~A.,  2017, \mn@doi [\mnras] {10.1093/mnras/stx759},
  \href {https://ui.adsabs.harvard.edu/abs/2017MNRAS.468.3985S} {468, 3985}

\bibitem[\protect\citeauthoryear{{Shultz}, {Wade}, {Rivinius}, {Neiner},
  {Henrichs}, {Marcolino}  \& {MiMeS Collaboration}}{{Shultz}
  et~al.}{2017}]{Shultz2017}
{Shultz} M.,  {Wade} G.~A.,  {Rivinius} T.,  {Neiner} C.,  {Henrichs} H.,
  {Marcolino} W.,   {MiMeS Collaboration} 2017, \mn@doi [\mnras]
  {10.1093/mnras/stx1632}, \href
  {https://ui.adsabs.harvard.edu/abs/2017MNRAS.471.2286S} {471, 2286}

\bibitem[\protect\citeauthoryear{{Shultz}, {Rivinius}, {Wade}, {Alecian},
  {Petit}  \& {BinaMIcS Collaboration}}{{Shultz} et~al.}{2018a}]{Shultz2018b}
{Shultz} M.,  {Rivinius} T.,  {Wade} G.~A.,  {Alecian} E.,  {Petit} V.,
  {BinaMIcS Collaboration} 2018a, \mn@doi [\mnras] {10.1093/mnras/stx3238},
  \href {https://ui.adsabs.harvard.edu/abs/2018MNRAS.475..839S} {475, 839}

\bibitem[\protect\citeauthoryear{{Shultz} et~al.,}{{Shultz}
  et~al.}{2018b}]{Shultz2018a}
{Shultz} M.~E.,  et~al., 2018b, \mn@doi [\mnras] {10.1093/mnras/sty103}, \href
  {https://ui.adsabs.harvard.edu/abs/2018MNRAS.475.5144S} {475, 5144}

\bibitem[\protect\citeauthoryear{{Shultz}, {Kochukhov}, {Wade}  \&
  {Rivinius}}{{Shultz} et~al.}{2018c}]{Shultz2018}
{Shultz} M.,  {Kochukhov} O.,  {Wade} G.~A.,   {Rivinius} T.,  2018c, \mn@doi
  [\mnras] {10.1093/mnrasl/sly070}, \href
  {https://ui.adsabs.harvard.edu/abs/2018MNRAS.478L..39S} {478, L39}

\bibitem[\protect\citeauthoryear{{Shultz}, {Rivinius}, {Das}, {Wade}  \& {Chand
  ra}}{{Shultz} et~al.}{2019a}]{Shultz2019c}
{Shultz} M.,  {Rivinius} T.,  {Das} B.,  {Wade} G.~A.,   {Chand ra} P.,  2019a,
  \mn@doi [\mnras] {10.1093/mnras/stz1129}, \href
  {https://ui.adsabs.harvard.edu/abs/2019MNRAS.486.5558S} {486, 5558}

\bibitem[\protect\citeauthoryear{{Shultz} et~al.,}{{Shultz}
  et~al.}{2019b}]{Shultz2019d}
{Shultz} M.~E.,  et~al., 2019b, \mn@doi [\mnras] {10.1093/mnras/stz2551}, \href
  {https://ui.adsabs.harvard.edu/abs/2019MNRAS.490..274S} {490, 274}

\bibitem[\protect\citeauthoryear{{Shultz} et~al.,}{{Shultz}
  et~al.}{2020}]{Shultz2020}
{Shultz} M.~E.,  et~al., 2020, \mn@doi [\mnras] {10.1093/mnras/staa3102}, \href
  {https://ui.adsabs.harvard.edu/abs/2020MNRAS.499.5379S} {499, 5379}

\bibitem[\protect\citeauthoryear{{Silvester} et~al.,}{{Silvester}
  et~al.}{2009}]{Silvester2009}
{Silvester} J.,  et~al., 2009, \mn@doi [\mnras]
  {10.1111/j.1365-2966.2009.15208.x}, \href
  {https://ui.adsabs.harvard.edu/abs/2009MNRAS.398.1505S} {398, 1505}

\bibitem[\protect\citeauthoryear{{St{\c{e}}pie{\'n}} \&
  {Landstreet}}{{St{\c{e}}pie{\'n}} \& {Landstreet}}{2002}]{Stepien2002}
{St{\c{e}}pie{\'n}} K.,  {Landstreet} J.~D.,  2002, \mn@doi [\aap]
  {10.1051/0004-6361:20020053}, \href
  {https://ui.adsabs.harvard.edu/abs/2002A&A...384..554S} {384, 554}

\bibitem[\protect\citeauthoryear{{Sundqvist}, {ud-Doula}, {Owocki}, {Townsend},
  {Howarth}  \& {Wade}}{{Sundqvist} et~al.}{2012}]{Sundqvist2012}
{Sundqvist} J.~O.,  {ud-Doula} A.,  {Owocki} S.~P.,  {Townsend} R. H.~D.,
  {Howarth} I.~D.,   {Wade} G.~A.,  2012, \mn@doi [\mnras]
  {10.1111/j.1745-3933.2012.01248.x}, \href
  {https://ui.adsabs.harvard.edu/abs/2012MNRAS.423L..21S} {423, L21}

\bibitem[\protect\citeauthoryear{{Takahashi} \& {Langer}}{{Takahashi} \&
  {Langer}}{2021}]{Takahashi2021}
{Takahashi} K.,  {Langer} N.,  2021, \mn@doi [\aap]
  {10.1051/0004-6361/202039253}, \href
  {https://ui.adsabs.harvard.edu/abs/2021A&A...646A..19T} {646, A19}

\bibitem[\protect\citeauthoryear{{Thompson} \& {Landstreet}}{{Thompson} \&
  {Landstreet}}{1985}]{Thompson1985}
{Thompson} I.~B.,  {Landstreet} J.~D.,  1985, \mn@doi [\apjl] {10.1086/184424},
  \href {https://ui.adsabs.harvard.edu/abs/1985ApJ...289L...9T} {289, L9}

\bibitem[\protect\citeauthoryear{{Wade}, {Donati}, {Landstreet}  \&
  {Shorlin}}{{Wade} et~al.}{2000}]{Wade2000}
{Wade} G.~A.,  {Donati} J.~F.,  {Landstreet} J.~D.,   {Shorlin} S.~L.~S.,
  2000, \mn@doi [\mnras] {10.1046/j.1365-8711.2000.03271.x}, \href
  {https://ui.adsabs.harvard.edu/abs/2000MNRAS.313..851W} {313, 851}

\bibitem[\protect\citeauthoryear{{Wade} et~al.,}{{Wade}
  et~al.}{2016}]{Wade2016}
{Wade} G.~A.,  et~al., 2016, \mn@doi [\mnras] {10.1093/mnras/stv2568}, \href
  {https://ui.adsabs.harvard.edu/abs/2016MNRAS.456....2W} {456, 2}

\bibitem[\protect\citeauthoryear{{Wade} et~al.,}{{Wade}
  et~al.}{2020}]{Wade2020}
{Wade} G.~A.,  et~al., 2020, \mn@doi [\mnras] {10.1093/mnras/staa025}, \href
  {https://ui.adsabs.harvard.edu/abs/2020MNRAS.492.2762W} {492, 2762}

\bibitem[\protect\citeauthoryear{{ud-Doula} \& {Owocki}}{{ud-Doula} \&
  {Owocki}}{2002}]{udDoula2002}
{ud-Doula} A.,  {Owocki} S.~P.,  2002, \mn@doi [\apj] {10.1086/341543}, \href
  {https://ui.adsabs.harvard.edu/abs/2002ApJ...576..413U} {576, 413}

\bibitem[\protect\citeauthoryear{{ud-Doula}, {Owocki}  \&
  {Townsend}}{{ud-Doula} et~al.}{2009}]{udDoula2009}
{ud-Doula} A.,  {Owocki} S.~P.,   {Townsend} R. H.~D.,  2009, \mn@doi [\mnras]
  {10.1111/j.1365-2966.2008.14134.x}, \href
  {https://ui.adsabs.harvard.edu/abs/2009MNRAS.392.1022U} {392, 1022}

\bibitem[\protect\citeauthoryear{{ud-Doula}, {Sundqvist}, {Owocki}, {Petit}  \&
  {Townsend}}{{ud-Doula} et~al.}{2013}]{udDoula2013}
{ud-Doula} A.,  {Sundqvist} J.~O.,  {Owocki} S.~P.,  {Petit} V.,   {Townsend}
  R.~H.~D.,  2013, \mn@doi [\mnras] {10.1093/mnras/sts246}, \href
  {https://ui.adsabs.harvard.edu/abs/2013MNRAS.428.2723U} {428, 2723}

\makeatother
\end{thebibliography}

\appendix
\newpage

\section{Radial velocity (RV) measurements from \esp~observations}

\begin{table}
\centering
\caption{Radial velocity (RV) measurements obtained from the intensity spectra (that is, from the individual subexposures of the  spectropolarimetric sequences) in 2020. RVs in 2019 were reported by \citet{Wade2020}.}
\label{table:rv_2020}
\begin{tabular}{c c c c}
\hline
\hline
HJD - 2450000 & RV (km s$^{-1}$) & HJD - 2450000 & RV (km s$^{-1}$) \\
\hline
9183.04882 & 7.8 $\pm$ 0.8 & 9190.06625 & 39.4 $\pm$ 1.4 \\
9183.04988 & 7.8 $\pm$ 0.8 & 9190.06732 & 39.4 $\pm$ 1.4 \\
9183.05095 & 7.8 $\pm$ 0.8 & 9190.96474 & 20.6 $\pm$ 1.1 \\
9183.05201 & 7.9 $\pm$ 0.8 & 9190.96581 & 19.9 $\pm$ 1.0 \\
9183.99882 & 39.2 $\pm$ 1.4 & 9190.96687 & 19.4 $\pm$ 1.0 \\
9183.99988 & 39.1 $\pm$ 1.4 & 9190.96794 & 19.1 $\pm$ 1.0 \\
9184.00094 & 38.9 $\pm$ 1.4 & 9191.93559 & 36.0 $\pm$ 1.4 \\
9184.00200 & 38.9 $\pm$ 1.4 & 9191.93665 & 36.3 $\pm$ 1.4 \\
9187.02426 & 7.7 $\pm$ 0.8 & 9191.93772 & 36.5 $\pm$ 1.4 \\
9187.02532 & 7.7 $\pm$ 0.8 & 9191.93878 & 36.8 $\pm$ 1.4 \\
9187.02638 & 7.6 $\pm$ 0.8 & 9192.02799 & 13.7 $\pm$ 0.9 \\
9187.02744 & 7.6 $\pm$ 0.8 & 9192.02905 & 13.3 $\pm$ 0.9 \\
9190.06412 & 39.2 $\pm$ 1.4 & 9192.03011 & 12.8 $\pm$ 0.9 \\
9190.06519 & 39.3 $\pm$ 1.4 & 9192.03118 & 12.6 $\pm$ 0.9 \\
\hline
\hline
\end{tabular}
\end{table}

\section{Table of Observations with HST/STIS}

\begin{table*}
\centering
\caption{STIS Spectroscopy of $\xi^1$ CMa (Program 14657, PI: Oskinova).}
\label{journal}
\begin{tabular}{ccccc}
\hline
\hline
ObsID & UT (Start) & ExpTime (s) & MJD (mid) & SNR\,$^{\ast}$ \\
\hline
od6n07010 & 2017-02-17T00:04:47 & 228 & 57801.0046 & 6.4 \\
od6n07020 & 2017-02-17T00:10:52 & 228 & 57801.0089 & 6.3 \\
od6n07030 & 2017-02-17T00:15:04 & 228 & 57801.0118 & 6.3 \\
od6n07040 & 2017-02-17T00:19:16 & 228 & 57801.0147 & 6.2 \\
od6n07050 & 2017-02-17T00:23:28 & 228 & 57801.0176 & 6.1 \\
od6n07060 & 2017-02-17T00:27:40 & 228 & 57801.0205 & 6.0 \\
od6n07070 & 2017-02-17T00:31:52 & 218 & 57801.0234 & 5.7 \\
od6n08010 & 2017-02-17T01:40:09 & 228 & 57801.0709 & 6.1 \\
od6n08020 & 2017-02-17T01:46:14 & 228 & 57801.0751 & 6.1 \\
od6n08030 & 2017-02-17T01:50:26 & 228 & 57801.0780 & 6.1 \\
od6n08040 & 2017-02-17T01:54:38 & 228 & 57801.0809 & 6.0 \\
od6n08050 & 2017-02-17T01:58:50 & 228 & 57801.0839 & 5.9 \\
od6n08060 & 2017-02-17T02:03:02 & 228 & 57801.0868 & 5.9 \\
od6n08070 & 2017-02-17T02:07:14 & 218 & 57801.0896 & 5.8 \\
od6n09010 & 2017-02-17T03:15:30 & 228 & 57801.1371 & 6.1 \\
od6n09020 & 2017-02-17T03:21:35 & 228 & 57801.1413 & 6.4 \\
od6n09030 & 2017-02-17T03:25:47 & 228 & 57801.1442 & 6.3 \\
od6n09040 & 2017-02-17T03:29:59 & 228 & 57801.1471 & 6.3 \\
od6n09050 & 2017-02-17T03:34:11 & 228 & 57801.1501 & 6.4 \\
od6n09060 & 2017-02-17T03:38:23 & 228 & 57801.1530 & 6.3 \\
od6n09070 & 2017-02-17T03:42:35 & 218 & 57801.1558 & 6.3 \\
od6n10010 & 2017-02-17T04:50:52 & 228 & 57801.2033 & 6.0 \\
od6n10020 & 2017-02-17T04:56:57 & 228 & 57801.2075 & 6.2 \\
od6n10030 & 2017-02-17T05:01:09 & 228 & 57801.2105 & 6.2 \\
od6n10040 & 2017-02-17T05:05:21 & 228 & 57801.2134 & 6.3 \\
od6n10050 & 2017-02-17T05:09:33 & 228 & 57801.2163 & 6.2 \\
od6n10060 & 2017-02-17T05:13:45 & 228 & 57801.2192 & 6.0 \\
od6n10070 & 2017-02-17T05:17:57 & 218 & 57801.2221 & 5.9 \\
od6n11010 & 2017-02-17T06:26:13 & 228 & 57801.2695 & 5.5 \\
od6n11020 & 2017-02-17T06:32:18 & 228 & 57801.2738 & 5.8 \\
od6n11030 & 2017-02-17T06:36:30 & 228 & 57801.2767 & 5.9 \\
od6n11040 & 2017-02-17T06:40:42 & 228 & 57801.2796 & 6.0 \\
od6n11050 & 2017-02-17T06:44:54 & 228 & 57801.2825 & 6.0 \\
od6n11060 & 2017-02-17T06:49:06 & 228 & 57801.2854 & 5.9 \\
od6n11070 & 2017-02-17T06:53:18 & 218 & 57801.2883 & 5.8 \\
od6n12010 & 2017-02-17T08:01:36 & 228 & 57801.3358 & 5.8 \\
od6n12020 & 2017-02-17T08:07:41 & 228 & 57801.3400 & 6.1 \\
od6n12030 & 2017-02-17T08:11:53 & 228 & 57801.3429 & 6.2 \\
od6n12040 & 2017-02-17T08:16:05 & 228 & 57801.3458 & 6.3 \\
od6n12050 & 2017-02-17T08:20:17 & 228 & 57801.3487 & 6.3 \\
od6n12060 & 2017-02-17T08:24:29 & 228 & 57801.3517 & 6.3 \\
od6n12070 & 2017-02-17T08:28:41 & 218 & 57801.3545 & 6.0 \\
\hline
\hline
\multicolumn{5}{l}{$^{\ast}$ Median value per pixel between 1493 and 1497~{\AA}.}\\
\end{tabular}
\end{table*}


\section{H\texorpdfstring{$\alpha$}{a} Equivalent Widths 
from \esp~Observations}

\begin{table*}
\centering
\caption{H$\alpha$ equivalent widths from the 2018, 2019, and 2020 \esp ~datasets.}
\label{table:appx_halpha}
\begin{tabular}{c c|c c|c c}
\hline
\hline
HJD-2458000 & EW (nm) & HJD-2458000 & EW (nm) & HJD-2458000 & EW (nm) \\  
\hline
8148.75321 & 0.0614 $\pm$ 0.0006 &	8557.84529 & 0.0586 $\pm$ 0.0006 &	8560.72580 & 0.0605 $\pm$ 0.0004	\\
8148.75429 & 0.0606 $\pm$ 0.0005 &	8557.84639 & 0.0592 $\pm$ 0.0004 &	8560.72690 & 0.0608 $\pm$ 0.0005	\\
8148.75537 & 0.0605 $\pm$ 0.0005 &	8557.84869 & 0.0578 $\pm$ 0.0005 &	8560.76130 & 0.0594 $\pm$ 0.0003 \\
8148.75645 & 0.0620 $\pm$ 0.0007 &	8557.84984 & 0.0576 $\pm$ 0.0005 &	8560.76242 & 0.0598 $\pm$ 0.0003	\\
8148.97144 & 0.0618 $\pm$ 0.0006 &	8557.85100 & 0.0583 $\pm$ 0.0005 &	8560.76352 & 0.0601 $\pm$ 0.0004	\\
8148.97252 & 0.0618 $\pm$ 0.0005 &	8557.85214 & 0.0589 $\pm$ 0.0004 &	8560.76463 & 0.0591 $\pm$ 0.0004	\\
8148.9736 & 0.06230 $\pm$ 0.0003 &	8559.76216 & 0.0569 $\pm$ 0.0004 &	8563.74274 & 0.0584 $\pm$ 0.0005	\\
8148.97468 & 0.0618 $\pm$ 0.0004 &	8559.76327 & 0.0567 $\pm$ 0.0005 &	8563.74385 & 0.0583 $\pm$ 0.0004	\\
8150.79889 & 0.0640 $\pm$ 0.0008 &	8559.76438 & 0.0581 $\pm$ 0.0004 &	8563.74496 & 0.0577 $\pm$ 0.0004	\\
8150.79997 & 0.0638 $\pm$ 0.0005 &	8559.76548 & 0.0562 $\pm$ 0.0004 &	8563.74607 & 0.0579 $\pm$ 0.0005	\\
8150.80105 & 0.0631 $\pm$ 0.0006 &	8559.76721 & 0.0574 $\pm$ 0.0005 &	8564.84618 & 0.0569 $\pm$ 0.0003	\\
8150.80213 & 0.0648 $\pm$ 0.0006 &	8559.76837 & 0.0569 $\pm$ 0.0003 &	8564.84728 & 0.0573 $\pm$ 0.0004 \\
8153.85512 & 0.0604 $\pm$ 0.0006 &	8559.76953 & 0.0570 $\pm$ 0.0004 &	8564.84839 & 0.0583 $\pm$ 0.0003	\\
8153.8562 & 0.06010 $\pm$ 0.0006 &	8559.77068 & 0.0566 $\pm$ 0.0005 &	8564.84949 & 0.0578 $\pm$ 0.0005	\\
8153.85728 & 0.0605 $\pm$ 0.0007 &	8559.77267 & 0.0570 $\pm$ 0.0004 &	8564.85109 & 0.0580 $\pm$ 0.0004	\\
8153.85835 & 0.0593 $\pm$ 0.0004 &	8559.78286 & 0.0566 $\pm$ 0.0005 &	8564.85220 & 0.0573 $\pm$ 0.0004	\\
8154.70789 & 0.0597 $\pm$ 0.0008 &	8559.78396 & 0.0559 $\pm$ 0.0004 &	8564.85331 & 0.0571 $\pm$ 0.0003	\\
8154.70898 & 0.0601 $\pm$ 0.0009 &	8559.78506 & 0.0564 $\pm$ 0.0004 &	9183.04882 & 0.0537 $\pm$ 0.0006	\\
8154.71008 & 0.0591 $\pm$ 0.0010 &	8559.78616 & 0.0562 $\pm$ 0.0004 &	9183.04988 & 0.0555 $\pm$ 0.0005	\\
8154.71119 & 0.0600 $\pm$ 0.0007 &	8559.78814 & 0.0566 $\pm$ 0.0005 &	9183.05095 & 0.0543 $\pm$ 0.0003	\\
8154.91334 & 0.0614 $\pm$ 0.0009 &	8559.78928 & 0.0563 $\pm$ 0.0006 &	9183.05201 & 0.0540 $\pm$ 0.0004	\\
8154.91444 & 0.0607 $\pm$ 0.0010 &	8559.79043 & 0.0563 $\pm$ 0.0005 &	9183.99882 & 0.0624 $\pm$ 0.0006	\\
8154.91554 & 0.0608 $\pm$ 0.0007 &	8559.79158 & 0.0572 $\pm$ 0.0005 &	9183.99988 & 0.0621 $\pm$ 0.0006	\\
8154.91664 & 0.0606 $\pm$ 0.0009 &	8559.79356 & 0.0563 $\pm$ 0.0005 &	9184.00094 & 0.0622 $\pm$ 0.0006	\\
8156.76397 & 0.0593 $\pm$ 0.0006 &	8559.79471 & 0.0562 $\pm$ 0.0004 &	9184.00200 & 0.0622 $\pm$ 0.0006	\\
8156.76505 & 0.0595 $\pm$ 0.0006 &	8559.79586 & 0.0569 $\pm$ 0.0005 &	9187.02426 & 0.0546 $\pm$ 0.0003	\\
8156.76613 & 0.0573 $\pm$ 0.0006 &	8559.79702 & 0.0568 $\pm$ 0.0005 &	9187.02532 & 0.0555 $\pm$ 0.0004	\\
8156.76721 & 0.0600 $\pm$ 0.0005 &	8559.81117 & 0.0560 $\pm$ 0.0004 &	9187.02638 & 0.0553 $\pm$ 0.0004	\\
8156.91427 & 0.0618 $\pm$ 0.0006 &	8559.81230 & 0.0563 $\pm$ 0.0005 &	9187.02744 & 0.0546 $\pm$ 0.0003	\\
8156.91535 & 0.0604 $\pm$ 0.0005 &	8559.81342 & 0.0557 $\pm$ 0.0004 &	9190.06412 & 0.0563 $\pm$ 0.0008	\\
8156.91643 & 0.0617 $\pm$ 0.0005 &	8559.81455 & 0.0554 $\pm$ 0.0005 &	9190.06519 & 0.0564 $\pm$ 0.0007	\\
8557.76676 & 0.0594 $\pm$ 0.0005 &	8559.82356 & 0.0561 $\pm$ 0.0004 &	9190.06625 & 0.0563 $\pm$ 0.0007	\\
8557.76787 & 0.0593 $\pm$ 0.0005 &	8559.82467 & 0.0559 $\pm$ 0.0003 &	9190.06732 & 0.0564 $\pm$ 0.0007	\\
8557.76897 & 0.0600 $\pm$ 0.0006 &	8559.82577 & 0.0571 $\pm$ 0.0003 &	9190.96474 & 0.0572 $\pm$ 0.0004	\\
8557.77007 & 0.0591 $\pm$ 0.0004 &	8559.82688 & 0.0563 $\pm$ 0.0003 &	9190.96581 & 0.0571 $\pm$ 0.0004	\\
8557.7724 & 0.05990 $\pm$ 0.0006 &	8560.71247 & 0.0603 $\pm$ 0.0005 &	9190.96687 & 0.0570 $\pm$ 0.0004	\\
8557.77349 & 0.0599 $\pm$ 0.0006 &	8560.71357 & 0.0607 $\pm$ 0.0006 &	9190.96794 & 0.0561 $\pm$ 0.0003	\\
8557.77459 & 0.0601 $\pm$ 0.0005 &	8560.71467 & 0.0603 $\pm$ 0.0004 &	9191.93559 & 0.0564 $\pm$ 0.0006	\\
8557.77569 & 0.0606 $\pm$ 0.0005 &	8560.71577 & 0.0612 $\pm$ 0.0004 &	9191.93665 & 0.0565 $\pm$ 0.0004	\\
8557.83743 & 0.0588 $\pm$ 0.0005 &	8560.71862 & 0.0615 $\pm$ 0.0004 &	9191.93772 & 0.0566 $\pm$ 0.0006	\\
8557.83856 & 0.0584 $\pm$ 0.0005 &	8560.71973 & 0.0605 $\pm$ 0.0004 &	9191.93878 & 0.0567 $\pm$ 0.0005	\\
8557.83969 & 0.0596 $\pm$ 0.0004 &	8560.72083 & 0.0598 $\pm$ 0.0004 &	9192.02799 & 0.0551 $\pm$ 0.0004	\\
8557.84081 & 0.0591 $\pm$ 0.0005 &	8560.72193 & 0.0603 $\pm$ 0.0004 &	9192.02905 & 0.0563 $\pm$ 0.0004	\\
8557.84309 & 0.0586 $\pm$ 0.0004 &	8560.72359 & 0.0611 $\pm$ 0.0004 &	9192.03011 & 0.0560 $\pm$ 0.0004	\\
8557.84419 & 0.0575 $\pm$ 0.0005 &	8560.72470 & 0.0606 $\pm$ 0.0003	\\	
\hline
\hline
\end{tabular}
\end{table*}


\bsp	
\label{lastpage}
\end{document}